\documentclass[aps,prb,showpacs,superscriptaddress,twocolumn,amssymb,amsfonts]{revtex4}
\usepackage{amsmath}
\usepackage{color}
\usepackage{bm}
\usepackage{epsfig}
\usepackage{graphics}
\usepackage{dsfont}
\usepackage[colorlinks,citecolor=blue]{hyperref}
\usepackage{multirow}
\usepackage{pdfsync}

\def\beq{\begin{equation}}
\def\eeq{\end{equation}}
\def\beqn{\begin{eqnarray}}
\def\eeqn{\end{eqnarray}}
\def\bpm{\begin{pmatrix}}
\def\epm{\end{pmatrix}}

\newcommand{\Ref}[1]{Ref.~\onlinecite{#1}}

\renewcommand{\bf}{\mathbf}

\newcommand{\imth}{\hspace{1pt}\mathrm{i}\hspace{1pt}}

\newcommand{\Rmnum}[1]{\expandafter\@slowromancap\romannumeral #1@}
\newcommand{\ie}{{\emph{i.e.~}}}
\newcommand{\eg}{{\emph{e.g.~}}}

\newcommand{\alert}[1]{{\color{red}{#1}}}
\newcommand{\bea}{\begin{eqnarray}}
\newcommand{\eea}{\end{eqnarray}}
\newcommand{\bal}{\begin{aligned}}
\newcommand{\eal}{\end{aligned}}

\newcommand{\bss}{{\boldsymbol{\sigma}}}
\newcommand{\bst}{{\boldsymbol{T}}}
\newcommand{\bse}{{\boldsymbol{e}}}
\newcommand{\cs}{{C_6}}
\newcommand{\rcs}{R_{\pi/3}}
\newcommand{\mbz}{{\mathbb{Z}}}
\newcommand{\bsr}{\emph{b}SR}
\newcommand{\fsr}{\emph{f}SR}

\newcommand{\dket}[1]{|{#1}\rangle}

\newcommand{\expval}[1]{\langle{#1}\rangle}

\newcommand{\ztsl}{$Z_2$ spin liquid}

\begin{document}
\title{Unification of bosonic and fermionic theories of spin liquids on the kagome lattice}
\author{Yuan-Ming Lu}
\affiliation{Department of Physics, University of California,
Berkeley, CA 94720} \affiliation{Materials Sciences Division,
Lawrence Berkeley National Laboratory, Berkeley, CA 94720}
\affiliation{Department of Physics, The Ohio State University, Columbus, OH 43210, USA}
\author{Gil Young Cho}
\affiliation{Department of Physics, University of California,
Berkeley, CA 94720}\affiliation{Department of Physics, Institute for Condensed Matter Theory,
University of Illinois at Urbana-Champaign, IL 61801}
\author{Ashvin Vishwanath}
\affiliation{Department of Physics, University of California,
Berkeley, CA 94720} \affiliation{Materials Sciences Division,
Lawrence Berkeley National Laboratory, Berkeley, CA 94720}
\date{\today}


%

\begin{abstract}
Recent numerical studies have provided strong evidence for a gapped $Z_2$ quantum spin liquid in the kagome lattice spin-1/2 Heisenberg model. A special feature of spin liquids is that symmetries can be fractionalized, and different patterns of symmetry fractionalization imply distinct phases. The symmetry fractionalization pattern for the kagome spin liquid remains to be determined. A popular approach to studying spin liquids is to decompose the physical spin into partons obeying either bose (Schwinger bosons) or fermi (Abrikosov fermions) statstics, which are then treated within the mean-field theory. A longstanding question has been whether these two approaches are truly distinct, or describe the same phase in complementary ways. Here we show that all 8 \ztsl~ phases in Schwinger-boson mean-field (SBMF) construction can also be described in terms of Abrikosov fermions, unifying pairs of theories that seem rather distinct. The key idea is that for \ztsl~states that admit a SBMF description on kagome lattice, the symmetry fractionalization of visions is uniquely fixed. Two promising candidate states for kagome Heisenberg model, Sachdev's $Q_{1}=Q_{2}$ SBMF state and Lu-Ran-Lee's $Z_2[0,\pi]\beta$ Abrikosov fermion state, are found to describe the same symmetric spin liquid phase. We expect these results to aid in a complete specification of the numerically observed spin liquid phase. We also discuss a set of \ztsl~phases in fermionic parton approach, where spin rotation and lattice symmetries protect gapless edge states, that do not admit a SBMF description.

\end{abstract}


\maketitle

%

\section{Introduction}\label{sec:intro}

$Z_2$ spin liquids (SLs) are a class of disordered many-spin states which have a finite energy gap for all bulk excitations. They differ fundamentally  from symmetry breaking ground states such as magnetically ordered phases and valence bond solids, since, in their simplest form,  they preserve all the symmetries including spin rotation, time reversal  and crystal symmetries. More importantly they possess bulk quasiparticles obeying fractional statistics\cite{Wilczek1990B}. For example in the most common $Z_2$ SL of a spin-1/2 system, there are three distinct types of fractionalized bulk excitations\cite{Read1991,Wen1991c,Jalabert1991,Mudry1994,Senthil2000,Kitaev2003}: \emph{bosonic spinon $b$} with half-integer spin, \emph{fermionic spinon $f$} with half-integer spin, and \emph{bosonic vison $v$} (a vortex excitation of $Z_2$ gauge theory) with integer spin. They all obey mutual semion statistics\cite{Kitaev2003}: \ie a bosonic spinon acquires a $-1$ Berry phase when it adiabatically encircles a fermionic spinon or a vison.  These statistical properties are identical to those of excitations in $Z_2$ gauge theory\cite{Kogut1979}, hence the name ``$Z_2$ spin liquid''.

Recently, interest in $Z_2$ SLs has been recharged by numerical studies on the spin-$1/2$ Heisenberg model on kagome\cite{Yan2011,Jiang2012a,Depenbrock2012,Suttner2014} lattice, where this state is strongly indicated. In particular a topological entanglement entropy\cite{Kitaev2006a,Levin2006} of $\gamma=\log2$ is observed in the ground state. Just like local order parameters used to describe symmetry breaking phases\cite{Landau1937}, here fractional statistics and topological entanglement entropy serve as fingerprints of the topological order\cite{Wen2004B} in $Z_2$ spin liquids. Analogous results have been reported for the frustrated square lattice, although the correlation lengths in that case are not as small as in the kagome lattice\cite{Wang2011a,Jiang2012,Hu2013a,Iqbal2013,Iqbal2014}.

Intriguingly, the experimentally studied spin-$1/2$ kagome materials - such as herbertsmithite\cite{Helton2007} - also remain quantum disordered down to the lowest temperature scales studied, well below the exchange energy scales. However, in contrast to the numerical studies, most experimental evidences point to a gapless ground state\cite{Savary2017,Norman2016}. It is currently still under debate if the gaplessness is an intrinsic feature\cite{Mendels2016} or a consequence of impurities that are known to be present in these materials\cite{Fu2015}. Furthermore the magnetic Hamiltonian of the material may depart from the pure Heisenberg limit. Relating the numerical results to experiments remains an important open question.

Since it preserves all symmetries of the system, is a $Z_2$ SL fully characterized by its topological order? The answer is no. In fact, the interplay of symmetry and topological order leads to a very rich structure. There are many different $Z_2$ spin liquids with the same $Z_2$ topological order and the same symmetry group, but they cannot be continuously connected to each other without breaking the symmetry: they are dubbed ``symmetry enriched topological (SET)'' phases\cite{Wen2002,Hung2013,Mesaros2013,Essin2013,Hu2013,Lu2016,Hung2013a,Barkeshli2014,Tarantino2016}. In a SET phase the quasiparticles not only have fractional statistics, but can also carry projective representation of the symmetry group. This phenomena is dubbed ``symmetry fractionalization''\cite{Essin2013,Barkeshli2014,Tarantino2016}, a well-known example being the fractional charge carried by the quasiholes (or quasielectrons) in the fractional quantum Hall effect\cite{Laughlin1983}. Different SET phases are characterized by different patterns of symmetry fractionalization, mathematically classified by the 2nd group cohomology\cite{Essin2013,Barkeshli2014,Tarantino2016} $\mathcal{H}^2(\mathcal{G}_s,\mathcal{A})$, where $\mathcal{G}_s$ is the symmetry group of the system and $\mathcal{A}$ is the (fusion) group of Abelian anyons in the topological order. In the case of $Z_2$ spin liquid, $\mathcal{A}=Z_2\times Z_2$ according the Abelian fusion rules summarized in (\ref{fusion rules}).

In the literature $Z_2$ SLs have been constructed in various slave-particle (or parton) frameworks: the most predominant two approaches fractionalize physical spin-1/2's into bosonic spinons\cite{Arovas1988,Read1991,Sachdev1992,Wang2006} and fermionic spinons\cite{Abrikosov1965,Baskaran1987,Affleck1988a,Affleck1988b,Wen1991c,Mudry1994,Wen2002} respectively. Both approaches yield variational wavefunctions with good energetics\cite{Tay2011,Ran2007,Iqbal2011} for the kagome lattice model. It was proposed that symmetric $Z_2$ SLs are classified by the projective symmetry groups\cite{Wen2002} (PSGs) of bosonic/fermionic spinons. However it has been a long-time puzzle to understand the relation between different PSGs in bosonic-spinon representation (\bsr) and fermionic-spinon representations (\fsr)\cite{Lu2011a}. To be specific in the kagome lattice Heisenberg model, in \bsr~(Schwinger-boson approach) there are 8 different $Z_2$ SLs\cite{Wang2006} among which the so-called $Q_1=Q_2$ state\cite{Sachdev1992} is considered a  promising candidate according to variational calculations\cite{Tay2011}. Meanwhile there are 20 distinct $Z_2$ SLs\cite{Lu2011} in \fsr~(Abrikosov-fermion approach), including the so-called $Z_2[0,\pi]\beta$ state\cite{Lu2011} which is in the neighborhood of energetically favorable $U(1)$ Dirac SL\cite{Ran2007}. Are these two candidate states actually two different descriptions of the same gapped phase? If not, what are their counterparts in the other representation?

In this paper we establish the general connection between different $Z_2$ SLs in \bsr~and \fsr. We show that $Z_2$ SLs constructed by projecting parton mean-field states in \bsr~(Schwinger-boson representation) cannot host symmetry-protected gapless edge states. This important observation allows us to determine how visons transform under symmetry in Schwinger-boson $Z_2$ SLs, and to further relate a Schwinger-boson state to an Abrikosov-fermion one. Since a bosonic spinon and vison fuse to a fermionic spinon as shown in (\ref{fusion rules}), their corresponding PSG coefficients naively should follow a product rule. However crucially, in some cases such as the PSG coefficients involving inversion symmetry\cite{Essin2013}, extra twist factors enter, modifying the naive fusion rule. Here we identify two additional instances where such nontrivial PSG fusion rules occur, as explained in Section \ref{sec:nontrivial fusion}. Related results can also be established using different techniques\cite{Zaletel2015,Qi2015}.

Next, we demonstrate that knowledge of just the bosonic (or just the fermionic) spinon PSG, with no further information such as the existence of a SBMF ansatz, is not enough to fully characterize a $Z_2$ SL. For example, two distinct $Z_2$ SLs in \fsr~can have the same PSG for fermionic spinons while only one of them has symmetry-protected gapless edge modes. However, they differ in the topology of spinon band structures, which provides an interesting link between symmetry implementation and topological edge states. By arguing the absence of symmetry-protected gapless edge states in any SBMF state, we show that all $Z_2$ spin liquids in SBMF construction must have a trivial PSG (or symmetry fractionalization pattern) for vison $v$. The knowledge of bosonic spinon PSG and vison PSG in a SBMF state leads to its fermionic spinon PSG, with the help of proper twist factors, therefore establishing the correspondence between a SBMF state and an Abrikosov-fermion $Z_2$ SL. Applying these general principles to $Z_2$ SLs on kagome lattice, we show that all 8 different Schwinger-boson (\bsr) states have their partners in the Abrikosov-fermion (\fsr) representation. In particular $Q_1=Q_2$ state\cite{Sachdev1992} in Schwinger boson representation belongs to the same SET phase as $Z_2[0,\pi]\beta$ state\cite{Lu2011} in Abrikosov fermion representation. This correspondence allows us to identify the possible symmetry-breaking phases in proximity to $Z_2$ SLs on kagome lattice. In fact all 8 SBMF states have their Abrikosov fermion counterparts, as summarized in Table \ref{tab:ansatz}). Part of these correspondences (for 4 SBMF states with $p_2+p_3=1$ in TABLE \ref{tab:ansatz}) has been obtained previously in \Ref{Yang2012}, by explicitly identifying their projected wavefunctions. These results serve as a useful guide in future studies of $Z_2$ SLs.

This article is organized as follows. After a brief review on symmetry fractionalization, PSG and their relations in $Z_2$ SLs in section \ref{sec:sym frac}-\ref{sec:psg}, we first establish the twist factors between PSGs of different anyons in a \ztsl~in section \ref{sec:nontrivial fusion}. In section \ref{sec:edge->vison} we show that absence of protected edge states and defect bound states in a \ztsl~can determine the vison PSG (see TABLE \ref{tab:unification}). These results allow us to compute the vison and fermion PSG in any SBMF state, which establishes to the correspondence between Schwinger-boson and Abrikosov-fermion mean-field states of \ztsl s, as studied in section \ref{UNIFICATION} and summarized in TABLE \ref{tab:ansatz}. In section \ref{sec:KHM candidate} we analyze and argue the most promising \ztsl~candidate for spin-1/2 kagome Heisenberg model, \ie $Q_1=Q_2$ SBMF state which is equivalent to the Abrikosov-fermion $Z_2[0,\pi]\beta$ state. Finally in section \ref{sec:tsc} we discuss possible \ztsl~s with mirror-symmetry protected edge states in the Abrikosov fermion representation.

\section{Symmetry fractionalization in a $Z_2$ spin liquid}\label{sec:general}

\subsection{A brief review on symmetry fractionalization}\label{sec:sym frac}

Symmetry fractionalization\cite{Essin2013,Barkeshli2014,Tarantino2016} is a mathematical framework that characterizes and classifies different SET phases in two spatial dimensions (2d). The key point is that a global (or crystalline) symmetry can act projectively on the anyons in a gapped 2d topological order. More precisely, the action $U_g^a$ of symmetry element $g$ on anyon $a$ satisfies the following condition in a gapped 2d topological order\footnote{We do not consider more complicated cases where symmetry $g$ can permute different anyons, which cannot happen in the symmetric spin-1/2 $Z_2$ spin liquids discussed in this work.}:
\bea\label{def:cohomology}
U_g^a\cdot U_h^a=\omega_a(g,h)~U^a_{gh}
\eea
where
\bea
\omega_a(g,h)=\langle\omega(g,h),a\rangle\in U(1)
\eea
is the $U(1)$-valued mutual braiding phase between $a$ and an Abelian anyon $\omega(g,h)$. This can be understood by considering defect $\tau_g$ of symmetry element $g$, which must satisfy the following fusion rule
\bea
\tau_g\times\tau_h=\tau_{gh}\times\omega(g,h)
\eea
where $\omega(g,h)$ can be an Abelian anyon in the topological order. These phase factors must satisfy the associativity condition
\bea\label{associativity:cocycle}
\omega_a(f,g)\omega_a(fg,h)=\omega_a(f,gh)\omega_a(g,h)
\eea
and be compatible with the fusion rules of anyons
\bea
\notag&a\times b=c\Longrightarrow \omega_a(g,h)\omega_b(g,h)=\omega_c(g,h),\\
&\text{if}~g,h~\text{are global (onsite) symmetries.}\label{psg:fusion rule:onsite}
\eea
Notice that there is a gauge redundancy for phase factors $\{\omega_a(g,h)\}$: we can always redefine the symmetry operation $U_g^a$ by adding an extra braiding phase $\langle a,\alpha_g\rangle\in U(1)$, where $\alpha_g$ is an arbitrary Abelian anyon. This gauge transformation modifies $\omega_a(g,h)$ by
\bea\label{gauge redundancy:cocycle}
\omega_a(g,h)\longrightarrow\omega_a(g,h)\frac{\langle a,\alpha_{gh}\rangle}{\langle a,\alpha_g\rangle\cdot\langle a,\alpha_h\rangle}
\eea

As a result, the gauge-inequivalent phase factors $\{\omega_a(g,h)\}$ are classified by the 2nd group cohomology ($\mathcal{G}_s$ denotes the symmetry group)
\bea
\{\omega_a(g,h)\}\in\mathcal{H}^2(\mathcal{G}_s,\mathcal{A})
\eea
with a discrete coefficient belonging to an Abelian group $\mathcal{A}$, \ie the (fusion) group of Abelian anyons in the topological order.

Take the $Z_2$ spin liquid for example, it features the following Abelian fusion rules\cite{Kitaev2003}:
\bea
&\notag b\times f=v,~~~b\times v=f,~~~f\times v=b,\\
&\label{fusion rules} b\times b=f\times f=v\times v=1.
\eea
where $b$ stands for the spin-1/2 bosonic spinon, $v$ for the spinless vison and $f$ for the spin-1/2 fermionic spinon. Here 1 stands for \emph{local} excitations carrying \emph{integer spins}, obeying the trivial bose statistics. From the above fusion rules, it is clear that all 3 types of Abelian anyons can be generated by 2 types of anyons among them, while the 3rd one can be obtained by fusing the other two anyons. These 2 ``elementary'' anyons can be chosen as any two types out of all three, such as $b$ and $v$. Since all anyons in (\ref{fusion rules}) satisfy a $Z_2$ fusion rule $a\times a=1$, this leads to an Abelian fusion group of $\mathcal{A}=Z_2\times Z_2$, where the two $Z_2$ factors are associated with $f$ and $v$ separately.

Now that phase factors $\{\omega_a(g,h)\}$ must be compatible with the $\mathcal{A}=Z_2\times Z_2$ fusion rules according to (\ref{psg:fusion rule:onsite}), since $\omega_1(g,h)\equiv1$ for an arbitrary local excitation $1$ we must have $\omega_a(g,h)=\pm1$ in a $Z_2$ spin liquid. Therefore the projective action of symmetry group $\mathcal{G}_s$ on the anyons in a $Z_2$ spin liquid is fully determined by
\bea
\notag&\{\omega_f(g,h)=\pm1|g,h\in \mathcal{G}_s\}\times\{\omega_v(g,h)=\pm1|g,h\in \mathcal{G}_s\}\\
&\in\mathcal{H}^2(\mathcal{G}_s,Z_2\times Z_2)\label{sym frac:z2 sl}
\eea
up to gauge redundancy. This completely characterizes the symmetry fractionalization pattern in a symmetric $Z_2$ spin liquid.

There is one more issue to emphasize: relation (\ref{psg:fusion rule:onsite}) from fusion rules only apply to global (``onsite'') symmetries. When elements $g$ or $h$ are crystalline symmetries, there can be an extra twist factors\cite{Essin2013,Zaletel2015} $\Omega_{a,b}^c(g,h)\in U(1)$ when we consider the implication of fusion rules on $\{\omega_a(g,h)\}$
\bea\label{fusion rule:psg}
a\times b=c\Longrightarrow\omega_a(g,h)\omega_b(g,h)=\Omega_{a,b}^c(g,h)\omega_c(g,h)
\eea
In the case of $Z_2$ spin liquid, as will be discussed soon in section \ref{sec:nontrivial fusion}, the $Z_2$-valued twist factors $\Omega_{a,b}^c(g,h)=\pm1$ can be nontrivial for cases involving crystalline rotation\cite{Essin2013}, mirror reflection and time reversal symmetries\cite{Zaletel2015}.

\subsection{Projective symmetry group (PSG) and its relation to symmetry fractionalization}\label{sec:psg}

Here we briefly review the concept of projective symmetry group (PSG) in the slave-particle construction of spin-1/2 quantum spin liquids\cite{Wen2002}. A more detailed discussion will be given later in section \ref{UNIFICATION}.

In the slave-particle construction, each spin-1/2 ${\bf S}_{\bf r}$ on lattice site ${\bf r}$ is represented by a Kramers pair of slave particles $\{\chi_{{\bf r},\alpha}|\alpha=\uparrow,\downarrow\}$ as
\bea
{\bf S}_{\bf r}=\frac12\sum_{\alpha,\beta=\uparrow,\downarrow}\chi^\dagger_{{\bf r},\alpha}\vec\sigma_{\alpha,\beta}\chi_{{\bf r},\beta}
\eea
where $\vec\sigma$ represents three Pauli matrices. The slave particles, or simply ``partons'', can obey either fermi or bose statistics: they correspond to the Abrikosov-fermion\cite{Abrikosov1965,Affleck1988b} (\fsr) or Schwinger-boson\cite{Auerbach1994B} (\bsr) representation respectively. Since the Hilbert space of partons are generally larger than the physical Hilbert space of spin-1/2, a ``single-occupancy'' constraint must be applied to the parton Hilbert space
\bea\label{single occupancy constraint}
\hat{n}_{\bf r}\equiv\sum_{\alpha=\uparrow,\downarrow}\chi^\dagger_{{\bf r},\alpha}\chi_{{\bf r},\alpha}=1,~~~\forall~{\bf r}.
\eea
As a result, the physical spin-1/2 wavefunction must be obtained by a Gutzwiller projection $\hat P_{\hat {n}_{\bf r}=1}$ on the parton wavefunction
\bea\label{gutz projection}
\dket{\Psi_{\text{spin}}}=\prod_{\bf r}\hat P_{\hat {n}_{\bf r}=1}\dket{\Psi_{\text{parton}}}.
\eea
or in other words
\bea
\expval{\alpha_1,\alpha_2,\cdots,\alpha_N|\Psi_{\text{spin}}}=\expval{0|\prod_{\bf r}\chi_{{\bf r},\alpha_{\bf r}}|\Psi_{\text{parton}}}
\eea

Physically, a symmetric spin-$1/2$ ground state must remain invariant under all symmetry operations $\{g\in \mathcal{G}_s\}$ of the symmetry group $\mathcal{G}_s$. On the other hand, is the parton state $\dket{\Psi_{\text{parton}}}$ also invariant under all symmetry operations? This is not necessarily true, because after the Gutzwiller projection (\ref{gutz projection}), any unitary rotations (``gauge rotations'') in the unphysical Hilbert space violating single-occupancy constraint (\ref{single occupancy constraint}) will not affect the physical spin-1/2 state $\dket{\Psi_{\text{spin}}}$. As a result, the (unprojected) parton state $\dket{\Psi_{\text{parton}}}$ only needs to remain invariant up to a gauge rotation. Therefore the symmetry action $U_g$ on the partons can be decomposed into the product of physical symmetry operation $O_g$ and gauge rotations ${G}_g$
\bea
U_g=O_g\cdot G_g,~~~\forall~g\in\mathcal{G}_s
\eea
Note that gauge rotations $\{G_g\}$ must all preserve the single-occupancy constraint (\ref{single occupancy constraint}) and the Gutzwiller-projected (physical) Hilbert space.

While the physical symmetry operation $O_g$ must form a linear representation of symmetry group $\mathcal{G}_s$ with
\bea\label{multiply:phys sym}
O_g\cdot O_h=O_{gh}
\eea
the gauge rotations generally form a projective representation of $\mathcal{G}_s$ as
\bea\label{multiply:gauge rot}
G_g\cdot G_h=\Lambda(g,h)~G_{gh}
\eea
where the parton state $\dket{\Psi_{\text{parton}}}$ must remain invariant under pure gauge rotation $\Lambda(g,h)$
\bea
\Lambda(g,h)\dket{\Psi_{\text{parton}}}=\dket{\Psi_{\text{parton}}},~~~\Lambda(g,h)\in\text{IGG}.
\eea
We define IGG (invariant gauge group)\cite{Wen2002} as the set of all gauge rotations that keep parton state $\dket{\Psi_{\text{parton}}}$ invariant. Clearly one can redefine the gauge rotation as $G_g\rightarrow G_g\cdot W_g$ by an extra IGG element $W_g\in~$IGG, leading to the following gauge redundancy on $\Lambda(g,h)$
\bea\label{gauge redundancy:psg}
\Lambda(g,h)\rightarrow \Lambda(g,h) W_g\cdot W_h\cdot W^{-1}_{gh},~~~\{W_g\}\in~\text{IGG}.
\eea
Due to associativity relations (\ref{multiply:phys sym}) and (\ref{multiply:gauge rot}), the symmetry actions $\{U_g\}$ on partons $\{\chi_{{\bf r},\alpha}\}$ also form a projective representation (or extension) of symmetry group $\mathcal{G}_s$:
\bea\label{def:psg}
U_g\cdot U_h=\Lambda(g,h)~U_{gh},~~~\Lambda(g,h)\in\text{IGG}.
\eea
$\{U_g|g\in\mathcal{G}_s\}$ is coined a ``projective symmetry group'' (PSG)\cite{Wen2002}, \ie an extension of symmetry group $\mathcal{G}_s$ satisfying
\bea
\text{PSG}/\text{IGG}=\mathcal{G}_s
\eea
It's straightforward to show that IGG elements $\{\Lambda(g,h)\}$ also satisfy the following associativity condition
\bea\label{associativity:psg}
\Lambda(f,gh)\Lambda(g,h)=\Lambda(f,g)\Lambda(fg,h)
\eea

In the slave-particle formalism, typically a mean-field Hamiltonian $\hat H_{MF}^{\phi}$ of partons will be constructed, giving rise to a parton ground state $\dket{\Psi_{\text{parton}}}$ (for details see section \ref{UNIFICATION}). In this case, IGG is the group of gauge rotations that keep the parton mean-field Hamiltonian invariant.

In the case of $Z_2$ spin liquids, there are both parton hopping and pairing terms in the mean-field Hamiltonian, and hence only the parton number parity is conserved in $\hat H_{MF}^{\phi}$ and $\dket{\Psi_{\text{parton}}}$. This leads to a $Z_2$ group structure of IGG~$=\{\chi_{{\bf r},\alpha}\rightarrow\pm\chi_{{\bf r},\alpha}\}\simeq Z_2$, generated by gauge rotation $W_0=(-1)^{\sum_{\bf r}\hat{n}_{\bf r}}$. In such a $Z_2$ spin liquid, PSG elements $\{U_g\}$ acts on partons in a projective fashion
\bea
\notag&U_gU_h\chi_{{\bf r},\alpha}U_h^{-1}U_g^{-1}=\eta(g,h)U_{gh}\chi_{{\bf r},\alpha}U_{gh}^{-1},\\
&\eta(g,h)=\pm1.
\eea
with IGG element $\Lambda(f,g)=\big[\eta(g,h)\big]^{\sum_{\bf r}\hat{n}_{\bf r}}$. A set of gauge invariant $Z_2$-valued phases $\{\eta(g,h)=\pm1|g,h\in\mathcal{G}_s\}$ therefore fully labels the PSG.

From associativity relation (\ref{associativity:cocycle}) and (\ref{associativity:psg}), gauge redundancy (\ref{gauge redundancy:cocycle}) and (\ref{gauge redundancy:psg}), and their definitions (\ref{def:cohomology}) and (\ref{def:psg}), the similarity between symmetry fractionalization pattern $\{\omega_a(g,h)\}$ and PSG pattern $\{\eta(g,h)\}$ is obvious. In fact, $Z_2$ PSGs are nothing but physical manifestions of abstract symmetry fractionalizations in gapped $Z_2$ spin liquids\cite{Essin2013}, where the anyon $a$ is determined by the statistics of partons $\{\chi_{{\bf r},\alpha}\}$. In other words, $\{\eta(g,h)=\pm1\}\simeq\{\omega_b(g,h)=\pm1\}$ in the Schwinger boson representation where $\chi\sim b$, and $\{\eta(g,h)=\pm1\}\simeq\{\omega_f(g,h)=\pm1\}$ in the Abrikosov fermion representation where $\chi\sim f$.

In the framework of symmetry fractionalization, relation (\ref{fusion rule:psg}) determined from fusion rules (\ref{fusion rules}) allows us to relate symmetry fractionalization patterns (or PSGs patterns in parton constructions) of all 3 anyon species $a=f,v,b$. Specifically, (\ref{fusion rule:psg}) allows us to determine the fermion PSG $\{\omega_f(g,h)\}$ from boson PSG $\{\omega_b(g,h)\}$ and vison PSG $\{\omega_v(g,h)\}$. This fact is crucial for the unification of Schwinger-boson and Abrikosov-fermion representations for gapped symmetric $Z_2$ spin liquids, as will become clear soon.

\begin{figure}
\includegraphics[width=1\columnwidth]{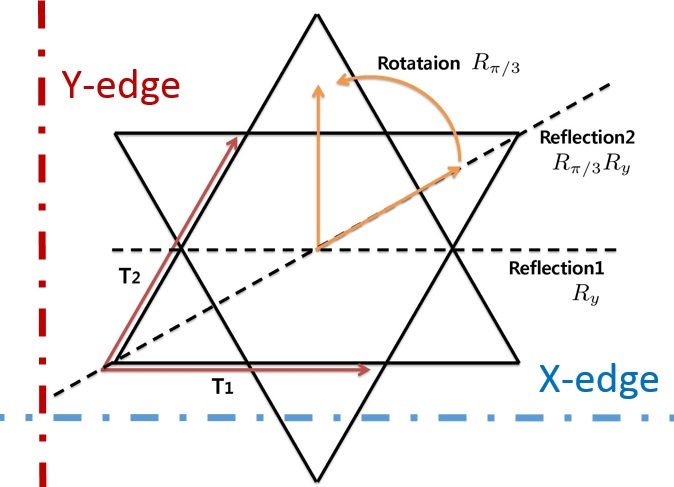}
\caption{Crystal symmetries of kagome lattice with 4 generators $\{T_{1,2},R_{\pi/3},R_y\}$. Translations ($T_1$, $T_2$) along the direction $1$ and $2$ are drawn as the directed arrow. $R_{\pi/3}$ stands for 60 degree rotation about a hexagon center. The mirror reflection 1 is denoted by $R_y$, while reflection 2 corresponds to $R_{\pi/3}R_y$.}
\label{RealUnitCell}
\end{figure}

\subsection{Twist factors for symmetry fractionalization in a \ztsl}\label{sec:nontrivial fusion}

In this section we establish the nontrivial twist factors for symmetry fractionalization patterns of different anyons $a=f,b,v$ in a \ztsl. As discussed earlier on fusion rules (\ref{fusion rules}), only two types of anyons are ``elementary'' in the sense that the 3rd type can be obtained by fusing these two types. Without loss of generality, here we choose fermionic spinon $f$ and vison $v$ as the two elementary anyons. Meanwhile the bosonic spinon $b$ is simply a bound state of $f$ and $v$, according to fusion rule $b=f\times v$. Focusing on 2d kagome lattice (space group $P6mm$, see FIG. \ref{RealUnitCell}) with symmetry group
\bea
\mathcal{G}_s=P6mm\times SO(3)\times Z_2^\bst
\eea
where $\bst$ refers to time reversal symmetry, we will reveal the nontrivial twist factors
\bea\label{twist factor}
\Omega_{f,v}^b(g,h)=\frac{\omega_f(g,h)\omega_v(g,h)}{\omega_b(g,h)}
\eea
These twist factors generally apply to an arbitrary 2d lattice.

Note that $\omega_{a}(g,h)$ for certain symmetry elements $g,h$ will depend on gauge choices due to redundancy (\ref{gauge redundancy:cocycle}). Therefore we need to focus on the gauge-invariant quantities, which sometimes can be a product of multiple $\{\omega_{a}(g,h)\}$ factors. Note that each PSG coefficient $\omega_a(g,h)$ is associated with algebraic identity $g\times h=gh$ for group elements $g,h$ and $gh$. Quite generally, the gauge-invariant coefficients can always be obtained from the following algebraic identity (summarized in the left column of TABLE \ref{tab:unification})
\bea
g_1\cdot g_2\cdots g_N=\bse,~~~\bse\equiv\text{identity element}.
\eea
as
\bea
\notag&\omega_a(g_1\cdots g_N)\equiv\\
&\omega_a(g_1,g_2)\cdot\omega_a(g_1g_2,g_3)\cdots\omega_a(\prod_{i=1}^{N-1}g_i,g_N)\label{def:psg coef}
\eea
Take the first algebraic identity in the left column of TABLE \ref{tab:ansatz} for example
\bea
T^{-1}_{2}T^{-1}_{1}T_{2}T_{1}=\bse
\eea
Its associated gauge-invariant phase factor is
\bea
\notag&\omega_a(T^{-1}_{2}T^{-1}_{1}T_{2}T_{1})\equiv\\
&=\omega_a(T^{-1}_{2},T^{-1}_{1})\cdot\omega_a(T_2,T_1)=\frac{\omega_a(T_2,T_1)}{\omega_a(T_1,T_2)}
\eea
As a result, the gauge-invariant twist factors are also associated with such algebraic identities
\bea
\notag&\Omega_{f,v}^b(g_1\cdots g_N)\equiv\\
&\omega_{f,v}^b(g_1,g_2)\cdot\omega_{f,v}^b(g_1g_2,g_3)\cdots\omega_{f,v}^b(\prod_{i=1}^{N-1}g_i,g_N)\label{def:twist factor}
\eea

In the following we discuss three different algebraic identities, each leading to one nontrivial twist factor in a \ztsl. We'll first present a general physical picture based on toric code model\cite{Kitaev2003} of $Z_2$ topological order, and then demonstrate them in projected parton wavefunctions for a \ztsl. We notice that aside from the twist factor $\Omega_{f,v}^b(I,I)=-1$ associated with algebraic identity $I^2=(R_{\pi/3})^6=\bse$, the other two nontrivial twist factors related to reflection symmetry $R_{x,y}$ and time reversal symmetry $\bst$ are missed in previous studies\cite{Essin2013}.\\

\subsubsection{Inversion $I$:~~~$\Omega_{f,v}^b(I,I)=-1$}\label{sec:nontrivial fusion:inversion square}

\begin{figure}
\includegraphics[width=1\columnwidth]{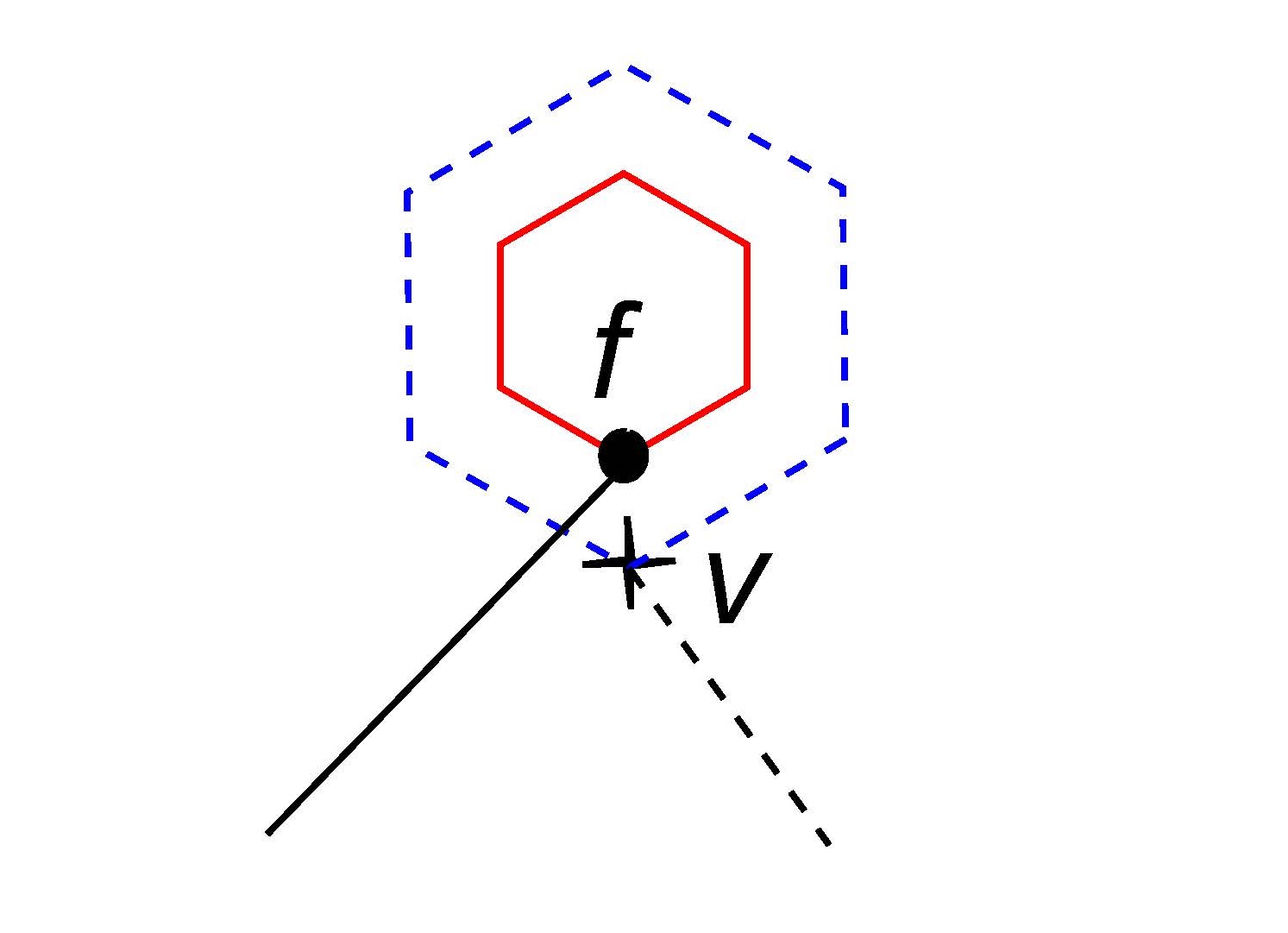}
\caption{(Color online) Nontrivial twist factor $\Omega_{f,v}^b(I,I)=-1$ associated with inversion square operation $(R_{\pi/3})^6=I^2=\bse$, as discussed in section \ref{sec:nontrivial fusion:inversion square}. The symmetry operator associated with $(R_{\pi/3})^6=I^2$ is illustrated by the solid (dashed) closed hexagon string with red (blue) color for fermionic spinons $f$ (vison $v$). The phase factor acquired by a bosonic spinon $b=f\times v$ in this process is $\omega_b(I,I)=-\omega_f(I,I)\cdot \omega_v(I,I)$, where the extra $-1$ sign comes from the crossing of fermion string (black solid line) and vison string (blue dashed line).}
\label{fig:inversion square}
\end{figure}

On the kagome lattice, hexagon-centered inversion symmetry operation $I=(R_{\pi/3})^3$ is the triple action of $\pi/3$ rotation $R_{\pi/3}$ (see FIG. \ref{RealUnitCell} and 8th row in TABLE \ref{tab:unification}). Clearly when inversion acts twice, all particles rotate counterclockwise around the hexagon center by a full circle \ie
\bea
I^2=(R_{\pi/3})^6=\bse.
\eea
Being a bound state of a vison $v$ and a fermionic spinon $f$, a bosonic spinon $b$ would collect an extra $-1$ phase factor\cite{Essin2013}, because the fermionic spinon encircles the vison once in this process.

To be more precise, let's introduce toric code model\cite{Kitaev2003} as a concrete demonstration of $Z_2$ topological orders (or $Z_2$ spin liquids). In the toric code model, ends of various open strings represent fractionalized excitations such as spinons and visons. There are three different types of strings, corresponding to three different anyons (bosonic spinon $b$, fermionic spinon $f$ and vison $v$) in a $Z_2$ spin liquid. In the figures we use solid lines to represent fermionic spinon (solid red circle) strings, and dashed lines for vison (blue cross) strings. Anyons of different types obey mutual semion statistics, which means each crossing of two different types of strings will yield a $-1$ phase factor.

As illustrated in FIG. \ref{fig:inversion square}, we consider a fermionic spinon $f$ on the end of a (black) solid string, and another vison $v$ on the end of a (black) dashed string. The bound state of these two object is a bosonic spinon $b=f\times v$. When the solid ($\hat S_f$) and dashed ($\hat S_v$) black string operators act on the ground state (vacuum) $\dket{\Psi_0}$, such an excited state
\bea
\dket{f\times v}=\hat S_f\hat S_v\dket{\Psi_0}
\eea
is created. As $\pi/3$ rotation acts for six times or equivalently inversion symmetry operation acts twice \ie $(R_{\pi/3})^6=I^2$, the phase factor acquired in this process is given by vacuum expectation value of the $I^2$ symmetry operator
\bea
\frac{\expval{a|(R_{\pi/3})^6|a}}{\expval{\Psi_0|(R_{\pi/3})^6|\Psi_0}}=\omega_a(I,I),~~~a=b,f,v.
\eea
which is the solid (dashed) closed hexagon string operator\cite{Essin2013} $\hat O_f$ ($\hat O_v$) with red (blue) color for fermionic spinon $f$ (vison $v$). More concretely we have
\bea
\notag&(R_{\pi/3})^6\hat S_f(R_{\pi/3})^{-6}=\hat O_f\cdot\hat S_f,\\
\notag&(R_{\pi/3})^6\hat S_v(R_{\pi/3})^{-6}=\hat O_f\cdot\hat S_v,\\
\notag&\frac{\expval{f\times v|(R_{\pi/3})^6|f\times v}}{\expval{\Psi_0|(R_{\pi/3})^6|\Psi_0}}=\expval{\Psi_0|S_v^{-1}S_f^{-1}O_fS_fO_vS_v|\Psi_0}\\
&=(S_vO_fS_v^{-1}O_f^{-1})\expval{\Psi_0|O_f|\Psi_0}\expval{\Psi_0|O_v|\Psi_0}\notag\\
&=-\frac{\expval{f|(R_{\pi/3})^6|f}}{\expval{\Psi_0|(R_{\pi/3})^6|\Psi_0}}\frac{\expval{v|(R_{\pi/3})^6|v}}{\expval{\Psi_0|(R_{\pi/3})^6|\Psi_0}}.
\eea
As the dashed blue string $\hat S_v$ crosses with the solid black string $\hat O_f$, an extra $-1$ sign will appear as we commute the $I^2$ symmetry operator for vison and the string operator for fermionic spinon. As a result the PSGs $\omega_a(I,I)$ associated with two inversion operations have a nontrivial twist factor $\Omega_{f,v}^b(I,I)=-1$. This conclusion remains true no matter the inversion symmetry is plaquette-centered or site-centered.\\

Now we confirm this intuitive picture using the projected parton wavefunction in the Abrikosov fermion representation. The same calculation also goes through in the Schwinger boson representation. Consider an excited state with a pair of fermionic spinons $f_{1,2}$ related by inversion symmetry $\hat I$
\bea
|f_{1,2}\rangle\equiv f_1 f_2|\Psi_0\rangle,~~~f_2=U_If_1U_I^{-1}.
\eea
where $|\Psi_0\rangle$ represents the parton ground state. By definition of symmetry fractionalization and PSGs we have
\bea
U_I^2 f_{1,2}U_I^{-2}=\omega_f(I,I)f_{1,2},~~~\omega_f(I,I)=\pm1.
\eea
It's straightforward to check that
\bea
\frac{\langle f_{1,2}|\hat{I}|f_{1,2}\rangle}{\langle\Psi_0|\hat{I}|\Psi_0\rangle}=-\omega_f(I,I)
\eea
where the extra $-1$ sign shows up because two fermionic spinon operators $f_1$ and $f_2$ are exchanged under inversion operation $\hat I$. Similarly for excited state
\bea
|v_{1,2}\rangle\equiv v_1v_2|\Psi_0\rangle,~~~U_Iv_1U_I^{-1}=v_2.
\eea
with a pair of visons $v_{1,2}$ on top of mean-field ground state, we have
\bea
\frac{\langle v_{1,2}|\hat{I}|v_{1,2}\rangle}{\langle\Psi_0|\hat{I}|Psi_0\rangle}=\omega_v(I,I),~~~U_I^2v_{1,2}U_I^{-2}=\omega_v(I,I)v_{1,2}
\eea
And the excited state with a pair of bosonic spinons $b_i=f_i\times v_i$ is created by
\bea
|b_{1,2}\rangle=f_1f_2v_1v_2|G\rangle.
\eea
Clearly we have
\bea
\frac{\langle b_{1,2}|\hat{I}|b_{1,2}\rangle}{\langle\Psi_0|\hat{I}|\Psi_0\rangle}=\omega_b(I,I)=-\omega_f(I,I)\cdot \omega_v(I,I)
\eea
And this proves the nontrivial twist factor
\bea
\Omega_{f,v}^b(I,I)\equiv\frac{\omega_f(I,I)\cdot \omega_v(I,I)}{\omega_b(I,I)}=-1
\eea
associated with algebraic identity $I^2=(R_{\pi/3})^6=\bse$.

\subsubsection{Mirror reflection $R$:~~~$\Omega_{f,v}^b(R,R)=-1$}\label{sec:nontrivial fusion:reflection square}

\begin{figure}
\includegraphics[width=1\columnwidth]{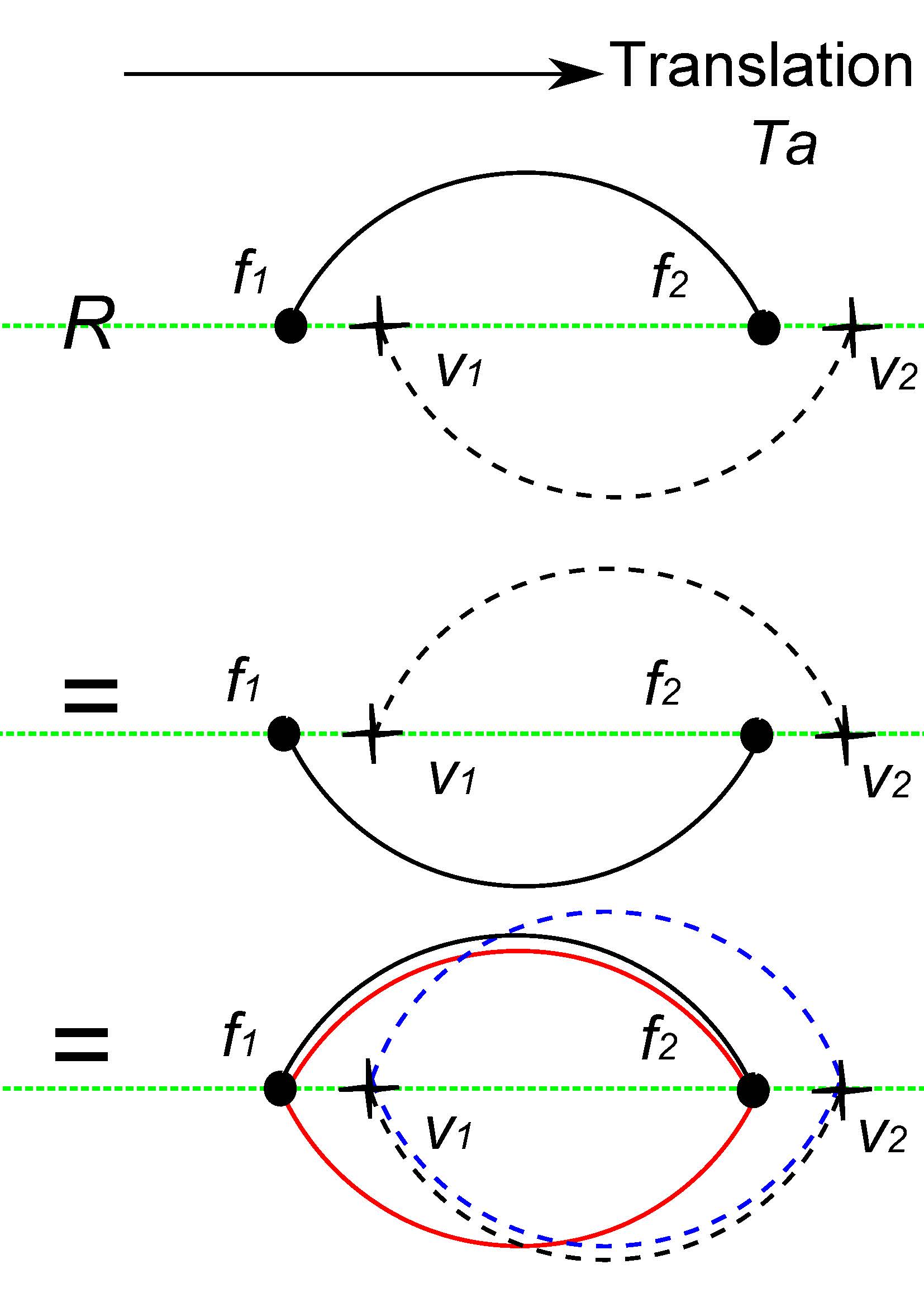}
\caption{(Color online) Nontrivial twist factor $\Omega_{f,v}^b(R,R)=-1$ associated with reflection square operation $R^2=\bse$, as discussed in section \ref{sec:nontrivial fusion:reflection square}. Reflection axis of $R$ is denoted by the dotted green line. The symmetry operator associated with reflection $R$ is illustrated by the solid (dashed) closed hexagon string $O_f(O_v)$ with red (blue) color for fermionic spinons $f_{1,2}$ (vison $v_{1,2}$), in the bottom of the figure. The phase factor acquired by each bosonic spinon $b_i=f_i\times v_i,~i=1,2$ in the process of $R^2=\bse$ is $\omega_b(R,R)=-\omega_f(R,R)\cdot \omega_v(R,R)$, where the extra $-1=S_fO_vS_f^{-1}O_v^{-1}$ sign comes from the crossing of open fermionic spinon string $S_f$ (black solid line) and closed vison string $O_v$ (blue dashed line). Note that the pair of fermionic spinons $f_{1,2}$ (and visons $v_{1,2}$) are related by translation $T_a$ (parallel to reflection axis), to guarantee that they share the same reflection quantum number.}
\label{fig:reflection square}
\end{figure}

In this case we consider a pair of fermionic spinons $f_{1,2}$ connected by a solid black string, and another pair of visons $v_{1,2}$ connected by a dashed black string as depicted in FIG. \ref{fig:reflection square}. We assume all these anyons lie on the reflection axis so that they are symmetric under reflection operation $R$. We'll reveal the nontrivial twist factor $\Omega_{f,v}^b(R,R)=-1$ associated with two reflection operations $R^2=\bse$, by studying the reflection quantum number carried by these anyons. A key ingredient of our discussions is that the pair of fermionic spinons $f_{1,2}$ (and visons $v_{1,2}$) must be related by translation $T_a$, as shown in FIG. \ref{fig:reflection square}. This guarantees two anyons of the same species share the same symmetry quantum number.

As illustrated in FIG. \ref{fig:reflection square}, the total reflection ($R$) quantum number for the pair of fermionic spinons $f_{1,2}$ (visons $v_{1,2}$) is the ground state expectation value of solid (dashed) closed string operator $\hat O_f$ ($\hat O_v$) with red (blue) color. Without loss of generality, we can assume this excited state with a pair of bosonic spinons is created by first applying the (solid black) fermionic spinon open string $\hat S_f$ on the ground state and then the (dashed black) vison open string $\hat S_v$:
\bea
&\dket{b_1b_2}=\hat S_v\cdot\hat S_f\dket{\Psi_0},\\
&\dket{f_1f_2}=\hat S_f\dket{\Psi_0},~~\dket{v_1v_2}=\hat S_v\dket{\Psi_0},\notag\\
&\notag U_R a_iU_R^{-1}=a_i,~~~a=b,f,v;~~i=1,2.
\eea
More concretely, in a \ztsl~we have (see FIG. \ref{fig:reflection square})
\bea
R\hat S_aR^{-1}=\hat O_a\cdot\hat S_a,~~~a=f,v.
\eea
as $(\hat S_a)^2=(\hat O_a)^2=1$ in a toric code. As a result we have
\bea
\frac{\expval{a_1a_2|R|a_1a_2}}{\expval{\Psi_0|R|\Psi_0}}=\expval{\Psi_0|\hat O_a|\Psi_0},~~a=f,v;
\eea
and
\bea
\notag&\frac{\expval{b_1b_2|R|b_1b_2}}{\expval{\Psi_0|R|\Psi_0}}=\frac{\expval{\Psi_0|S_f S_v(O_vS_v)(O_fS_f)R|\Psi_0}}{\expval{\Psi_0|R|\Psi_0}}\\
&=\expval{\Psi_0|\hat O_f|\Psi_0}\expval{\Psi_0|\hat O_v|\Psi_0}\cdot(S_fO_vS_f^{-1}O_v^{-1})\notag\\
&=-\frac{\expval{f_1f_2|R|f_1f_2}}{\expval{\Psi_0|R|\Psi_0}}\frac{\expval{v_1v_2|R|v_1v_2}}{\expval{\Psi_0|R|\Psi_0}}.
\eea

Since the dashed blue (closed) string anti-commute with black solid string, the reflection quantum number of bosonic spinon pair in FIG. \ref{fig:reflection square} equals the reflection quantum number of fermionic spinon pair multiplying that of vison pair, with an extra $-1$ sign. Since the pair of anyons are related by translation symmetry, this $-1$ sign should be evenly split into two halves: \ie upon reflection operation $R$, each bosonic spinon acquires an extra $\pm\imth$ phase in addition to the product of the fermionic spinon and vison. As a result when reflection $R$ acts twice, the phase factor acquired by each bosonic spinon $b_i=f_i\times v_i,~~i=1,2$ in this process is given by
\bea\notag
&\notag\omega_b(R,R)=(\pm\imth)^2\cdot \omega_f(R,R)\cdot \omega_v(R,R)\\
&=-\omega_f(R,R)\cdot \omega_v(R,R)
\eea
Therefore the twist factor associated with two reflection operations $R^2=\bse$ is indeed nontrival
\bea
\Omega_{f,v}^b(R,R)\equiv\frac{\omega_f(R,R)\cdot \omega_v(R,R)}{\omega_b(R,R)}=-1.
\eea

To validate this conclusion for reflection square $R^2=\bse$ in the Abrikosov fermion representation, we can follow exactly the same calculations as in the preceding section for inversion square $I^2=\bse$, by simply replacing inversion $I$ with reflection $R$.\\

\subsubsection{Time reversal $T$ and mirror reflection $R$: ${\Omega_{f,v}^b(R,T)}=-\Omega_{f,v}^b(T,R)$}\label{sec:nontrivial fusion:tr+ref}

Unlike other global (on-site) unitary symmetries, time reversal is an anti-unitary symmetry operation involving a complex conjugation. Below we show a nontrivial twist factor
\bea
\Omega_{f,v}^b(R^{-1}T^{-1}RT)=\frac{\Omega_{f,v}^b(R,T)}{\Omega_{f,v}^b(T,R)}=-1
\eea
associated with algebraic identity $R^{-1}T^{-1}RT=\bse$.

Let's first consider the combination $T\cdot R$ of time reversal $T$ and mirror reflection $R$, which is an anti-unitary spatial symmetry. When this symmetry acts twice, each anyon $a$ acquires a phase factor $\omega_a(TR,TR)=\pm1$. In this case, is there a nontrivial twist factor
\bea\label{twist:(TR)^2}
\Omega_{f,v}^b(TR,TR)=\frac{\omega_f(TR,TR)\omega_v(TR,TR)}{\omega_b(TR,TR)}
\eea
when fusing different anyons or not? The answer is no. In the same setup as in FIG. \ref{fig:reflection square}, each bosonic spinon gets an extra $\pm\imth$ phase (in addition to the phase factors acquired by fermionic spinon and vison) under reflection $R$. However when time reversal acts, it takes complex conjugation and hence we have $(\pm\imth)^\ast(\pm\imth)=1$. This extra phase hence cancels out upon symmetry operation $(TR)^2=\bse$. Therefore twist factor (\ref{twist:(TR)^2}) associated with $(TR)^2=\bse$ is trivial and equals unity. Meanwhile, as an algebraic identity we have
\bea
(TR)^2=(R^{-1}T^{-1}RT)\cdot (T^2)\cdot(R^2)=\bse.
\eea
and therefore from associativity we have
\bea
&\notag\Omega_{f,v}^b(R^{-1}T^{-1}RT)\equiv\frac{\Omega_{f,v}^b(R,T)}{\Omega_{f,v}^b(T,R)}\\
&=\frac{\Omega_{f,v}^b(TR,TR)}{\Omega_{f,v}^b(T,T)\Omega_{f,v}^b(R,R)}=-1\label{twist:T R commutator}
\eea
Note that for global time reversal symmetry, the twist factor is trivial \ie $\Omega_{f,v}^b(T,T)=+1$.

The above twist factor $\Omega_{f,v}^b(TR,TR)=1$ can also be argued in an alternative way. It is well-known that any excitation (labeled as $a$ here) in a time reversal invariant system can be categorized into Kramers doublets with $\omega_a(T,T)=-1$, and Kramers singlets with $\omega_a(T,T)=1$. In particular for each excitation $a$ with $\omega_a(T,T)$, there is a two-fold degeneracy (Kramers ``doublet'') protected by time reversal symmetry $T$. The Kramers doublets obey a $Z_2$ fusion rule, in the sense that the bound state of two Kramers doublets becomes one Kramers singlet. This implies a trivial twist factor $\Omega_{f,v}^b(T,T)=1$ for $T^2=\bse$ associated withe time reversal symmetry.

Now we consider an excitation $d$ located on the reflection plane, hence invariant under mirror reflection. Now that $TR$ is also an anti-unitary $Z_2$ symmetry just like $T$, for similar reasons there are also ``non-Kramers doublets'' with $\omega_d(TR,TR)=-1$, which features two-fold degeneracy protected by symmetry $TR$. From the viewpoint of this excitation $d$ on reflection plane, $TR$ symmetry can be treated in exactly the same fashion as time reversal $T$. Therefore non-Kramers doublets of anti-unitary $Z_2$ symmetry $TR$ should also obey a $Z_2$ fusion rule, \ie two non=Kramers doublets fuse into a non-Kramers singlet $s$ with $\omega_s(TR,TR)=1$. If we consider one fermonic spinon $f$ and a vison $v$, both on the reflection plane, it is straightforward to verify the trivial twist factor
\bea\label{twist:(TR)^2=+1}
\Omega_{f,v}^b(TR,TR)=\frac{\omega_f(TR,TR)\omega_v(TR,TR)}{\omega_b(TR,TR)}=1
\eea
based on the $Z_2$ fusion rules of non-Kramers doublets.\\

To summarized, we have established the nontrivial twist factor (\ref{twist:T R commutator}) associated with algebraic identity $R^{-1}T^{-1}RT=\bse$ using two different arguments. Notice that inversion symmetry $I$ is a combination of two reflection symmetries with perpendicular reflection planes \ie $I=R_xR_y$, as shown in FIG. \ref{RealUnitCell}. Therefore the twist factor associated with $I^{-1}T^{-1}IT$ is trivial
\bea
\Omega_{f,v}^b(I^{-1}T^{-1}IT)\equiv\frac{\Omega_{f,v}^b(I,T)}{\Omega_{f,v}^b(T,I)}=+1\label{twist:T I commutator}
\eea

Before closing of this section, we want to mention that all arguments used here can be made more rigorous by considering a thin cylinder geometry, which relates the phase factors $\{\omega_a(g,h)\}$ to one-dimensional invariants of symmetry protected topological (SPT) phases\cite{Chen2013}. This dimensional reduction approach is discussed in \Ref{Zaletel2015}.

\begin{table*}[tb]
\begin{tabular} {|c|c|c|c|c|c|c|}
\hline
Algebraic Identities&  bosonic spinon $b_{\sigma}$ & fermionic spinon $f_{\sigma}$ & vison $v$&Phase factors (\ref{def:psg coef})&Twist factors (\ref{def:twist factor})\\ \hline
$T^{-1}_{2}T^{-1}_{1}T_{2}T_{1}=\bse$&(-1)$^{p_1}$&$\eta_{12}$&-1&${\omega_a(T_2,T_1)}/{\omega_a(T_1,T_2)}$&1\\ \hline
$T^{-1}_{1}R^{-1}_{\pi/3}T_{2}R_{\pi/3}=\bse$&1& 1&1&+1 by gauge choice&1\\ \hline
$T^{-1}_{1}T_{2}R^{-1}_{\pi/3}T_{1}R_{\pi/3}=\bse$&1&1&1&+1 by gauge choice&1\\ \hline
$T_{1}R^{-1}_{x}T_{1}R_{x}=\bse$&1&1&1&${\omega_a(T_1R_x,T_1R_x)}/{\omega_a(R_x,R_x)}$&1\\ \hline
$T_{y}R^{-1}_{y}T_{y}R_{y}=\bse$&1&1&1&${\omega_a(T_yR_y,T_yR_y)}/{\omega_a(R_y,R_y)}$&1\\ \hline
$R_x^2\equiv(R_{\pi/3}R_{y})^2=\bse$&(-1)$^{p_2+p_3}$&$\eta_\bss$&1&$\omega_a(R_x,R_x)$&-1\\ \hline
$(R_{y})^{2}=\bse$&(-1)$^{p_2}$&$\eta_\bss\eta_{\bss\cs}$&1&$\omega_a(R_y,R_y)$&-1\\ \hline
$(R_{\pi/3})^{6}=I^2=\bse$&(-1)$^{p_1+p_3}$&$\eta_\cs$&1&$\omega_a(I,I)$&-1\\ \hline
$T_1^{-1}T^{-1}T_1T=\bse$&1&1&1&${\omega_a(T_1,T)}/{\omega_a(T,T_1)}$&1\\ \hline
$T_2^{-1}T^{-1}T_2T=\bse$&1&1&1&${\omega_a(T_2,T)}/{\omega_a(T,T_2)}$&1\\ \hline
$R_y^{-1}T^{-1}R_yT=\bse$&(-1)$^{p_2}$&$\eta_{\bss\bst}\eta_{\cs\bst}$&1&${\omega_a(R_y,T)}/{\omega_a(T,R_y)}$&-1\\ \hline
$R_x^{-1}T^{-1}R_xT=\bse$&(-1)$^{p_2+p_3}$&$\eta_{\bss\bst}$&1&${\omega_a(R_x,T)}/{\omega_a(T,R_x)}$&-1\\ \hline
$T^2=\bse$&-1&-1&1&$\omega_a(T,T)$&1\\ \hline
\end{tabular}
\caption{Algebraic identities $g_1\cdots g_N=\bse$, their associated phase factors (or PSG coefficients) $\omega_a(g_1\cdots g_N)$ in (\ref{def:psg coef}), and twist factors $\Omega_{f,v}^b(g_1\cdots g_N)$ in (\ref{def:twist factor}). Considering spin-1/2 \ztsl s on the kagome lattice, we list $\pm1$-valued PSG coefficients (phase factors) $\{\omega_a(g_1\cdots g_N)\}$ for bosonic spinon\cite{Wang2006} with $a=b$, fermionic spinon\cite{Lu2011} with $a=f$ and vison\cite{Huh2011} with $a=v$. Mirror reflection $R_x$ is defined as $R_x\equiv(R_{\pi/3})^3R_y$ and inversion $I\equiv(R_{\pi/3})^3$, see FIG. \ref{RealUnitCell}. We also introduce the translation $T_y\equiv T_1^{-1}T_2^2$ along $\hat y$-axis. Here bosonic spinon ($b_\sigma$) PSGs are labeled by three $Z_2$ integers\cite{Wang2006} $p_i=0,1~(i=1,2,3)$, while fermionic spinon ($f_\sigma$) PSGs are labeled by six signs\cite{Lu2011} $(\eta_{12},\eta_\bss,\eta_{\bss\cs},\eta_{\cs},\eta_{\bss\bst},\eta_{\cs\bst})$ where $\eta=\pm1$. Choosing a proper gauge we can always fix $\omega_a(T^{-1}_{1}R^{-1}_{\pi/3}T_{2}R_{\pi/3})=\omega_a(T^{-1}_{1}T_{2}R^{-1}_{\pi/3}T_{1}R_{\pi/3})\equiv1$ for all anyons $\{a=b,f,v\}$. Note that the vison PSG in \emph{any} SBMF $Z_2$ SL (\ie in \bsr)~is completely fixed as summarized above. If one \ztsl~state in \bsr~and one in \fsr~belong to the same SET phase, according to fusion rule condition (\ref{fusion rule:psg}), their PSG coefficients must satisfy relation (\ref{psg condition}).}
\label{tab:unification}
\end{table*}

\section{Implications on vison PSGs from absence of edge states}\label{sec:edge->vison}

In previous sections, we introduced symmetry fractionalization patterns $\{\omega_a(g,h)|g,h\in\mathcal{G}_s;~a=f,v\}$ to characterize a gapped symmetric \ztsl. In particular, the PSG coefficients $\{\omega_f(g,h)\}$ ($\{\omega_b(g,h)\}$) in the Abrikosov fermion (Schwinger boson) representation are nothing but symmetry fractionalization patterns for fermionic (bosonic) spinons $f$ ($b$). In other words, in each parton construction, the parton PSG only specifies the symmetry fractionalization pattern for one anyon species ($f$ or $b$). Meanwhile by establishing the nontrivial twist factors (\ref{twist factor}) in section \ref{sec:nontrivial fusion}, we can relate fermionic spinon PSG $\{\omega_f(g,h)\}$ and bosonic spinon PSG $\{\omega_b(g,h)\}$ via the vison PSG $\{\omega_v(g,h)\}$. In other words, knowing the vison PSG $\{\omega_v(g,h)\}$ in a Schwinger boson mean-field (SBMF) state will allow us to also compute the fermionic spinon PSG $\{\omega_f(g,h)\}$ in this state. By comparing it with the Abrikosov fermion \ztsl~states, we can establish the correspondence between this SBMF state and another possible Abrikosov fermion state. 

In this section, we answer the following question: what are the physical manifestations of the vison PSG in a \ztsl? As will become clear later in section \ref{UNIFICATION}, understanding this question will allow us to fix the vison PSGs in \emph{all} SBMF \ztsl~states, hence establishing the correspondence between Schwinger boson and Abrikosov fermion representations of \ztsl~states.

An important measurable property of topological phases are their edge states. Although $Z_2$ SLs in the absence of symmetries are expected to have gapped edges\cite{Kitaev2012}, the edge may be gapless due to the protection of certain symmetries\cite{Kou2009,Cho2012,Lu2016}; or in the case of discrete symmetries, spontaneously break symmetry on the edge. In particular, a symmetrically gapped edge of spin-1/2 $Z_2$ SLs with $SU(2)$ spin rotation symmetry puts strong constraints on the vison PSG, as will be argued below. 

The edge modes of a $Z_2$ SL can be fermionized\cite{Kou2008} with the same number of right ($\psi_{R,\alpha}$) and left ($\psi_{L,\alpha}$) movers (velocity is set to unity):
\bea
\notag&\mathcal{L}_0=\sum_{\alpha}\imth\psi^\dagger_{R,\alpha}(\partial_t-\partial_x)\psi_{R,\alpha}-\imth\psi^\dagger_{L,\alpha}(\partial_t+\partial_x)\psi_{L,\alpha}\\
&
\eea
where $\alpha$ denotes different branches of left/right movers. One can always add backscattering terms to gap out the edge modes $\{\psi_{R/L,\alpha}\}$
\bea
\mathcal{L}_1=\sum_{\alpha,b}\psi^\dagger_{R,\alpha}{\bf M}_{\alpha,\beta}\psi_{L,\beta}+\psi_{R,\alpha}{\mathbf{\Delta}}_{\alpha,\beta}\psi_{L,\beta}+~h.c.
\eea
if they are not forbidden by symmetry. In a different language, the above ``mass'' terms correspond to Bose condensation of either bosonic spinons $b$ or visons $v$ on the edge\cite{Bravyi1998,Kitaev2012} of a $Z_2$ SL. Since the bosonic spinons carry spin-$1/2$ each, condensing them will necessarily break spin rotational symmetry on the edge. Therefore the only way to obtain a gapped edge without breaking the symmetry is to condense visons, unless their symmetry fractionalization pattern (PSG) $\{\omega_v(g,h)\}$ will forbid it. In particular, if the symmetries (preserved on the edge) act projectively on the visons, it is impossible to condense visons without breaking the symmetries. The relevant symmetries here are the ones that leave the physical edge unchanged, \eg at least one translation symmetry among $T_{1,2}$ will be broken by the edge. Therefore the absence of symmetry protected edge states provides a strong constraint on the vison PSG.

Take kagome lattice for instance, on a cylinder with open boundaries parallel to $T_1$ direction (X-edge in FIG. \ref{RealUnitCell}), the remaining symmetries are translation $T_1$, time reversal $T$ and mirror reflection $R_x\equiv(R_{\pi/3})^3R_y$. If there are no symmetry protected edge states, then the remaining symmetries must act trivially (\ie not projectively) on visons:
\bea
\notag&T_1^{-1}T^{-1}T_1T=\bse\longrightarrow\frac{\omega_v(T_1,T)}{\omega_v(T,T_1)}=1,\\
\notag&R_x^{-1}T^{-1}R_xT=\bse\longrightarrow\frac{\omega_v(R_x,T)}{\omega_v(T,R_x)}=1,\\
&\notag R_x^2=\bse\longrightarrow{\omega_v(R_x,R_x)}=1,\\
&T_1R_x^{-1}T_1R_x=\bse\longrightarrow\frac{\omega_v(T_1R_x,T_1R_x)}{\omega_v(R_x,R_x)}=1.\label{PSG:T1 no edge}
\eea
so that no backscattering term is forbidden by symmetry.

Another inequivalent edge is perpendicular to $T_1$ direction (Y-edge in FIG. \ref{RealUnitCell}), which preserves translation
\bea
T_y\equiv T_1^{-1}T_2^2\notag
\eea
time reversal $T$ and mirror reflection $R_y$. Similarly, absence of protected edge modes necessarily implies that 
\bea
\notag&T_y^{-1}T^{-1}T_yT=\bse\longrightarrow\frac{\omega_v(T_y,T)}{\omega_v(T,T_y)}=1,\\
\notag&R_y^{-1}T^{-1}R_yT=\bse\longrightarrow\frac{\omega_v(R_y,T)}{\omega_v(T,R_y)}=1,\\
&\notag R_y^2=\bse\longrightarrow{\omega_v(R_y,R_y)}=1,\\
&T_yR_y^{-1}T_yR_y=\bse\longrightarrow\frac{\omega_v(T_yR_y,T_yR_y)}{\omega_v(R_y,R_y)}=1.\label{PSG:T1^(-1)T2^2 no edge}
\eea
\ie symmetry operations on visons form a linear (not projective) representation of the edge symmetry group.

As summarized in TABLE \ref{tab:unification}, by choosing all possible edge configurations, one can fix all gauge-invariant vison PSG coefficients in TABLE \ref{tab:unification} (for detailed derivations see Appendix \ref{app:vison PSG}) except for the three coefficients below
\bea
\notag\frac{\omega_v(T_2,T_1)}{\omega_v(T_1,T_2)},~~~\omega_v(T,T),~~~\omega_v(I,I).
\eea
In a Mott insulator (with an odd number of spin 1/2 moments per unit cell) like the spin-1/2 kagome lattice, any unfractionalized featureless phase with a spin gap must double the unit cell. Therefore the vision PSG for translations $T_{1,2}$ must satisfy:
\beq
T^{-1}_{2}T^{-1}_{1}T_{2}T_{1}=\bse\longrightarrow\frac{\omega_v(T_2,T_1)}{\omega_v(T_1,T_2)}=-1
\eeq
This may be thought of as visons acquiring an extra $-1$ phase factor on going around a unit cell containing an odd number of spinons, under the operations of $T^{-1}_{2}T^{-1}_{1}T_{2}T_{1}$. 

Meanwhile, since visons are spinless particles, they transform trivially as a Kramers singlet under time reversal $T$ and hence
\bea
T^2=\bse\longrightarrow\omega_v(T,T)=1.
\eea

Finally we present one argument to fix $\omega_v(I,I)$ associated with algebraic identity $I^2=(R_{\pi/3})^6=\bse$. In addition to edge states, symmetry-protected in-gap bound state in crystal defects\cite{Kroner1975,Kleman2008} (such as dislocations and disclinations) is another signature to probe crystal symmetry fractionalization. Just like gapless edge modes, the absence of defect bound states implies a trivial vison PSG of the associated crystal symmetry. In our case of kagome lattice, the crystal defect associated with $R_{\pi/3}$ rotation is a disclination, centered on the hexagon with Frank angle $\Omega=n\pi/3,~n\in\mathds{Z}$. Now consider a vison encircling around an elementary disclination with $\Omega=\pi/3$ for six times counterclockwise, and the phase factor it picks up in this process equals to the vison PSG coefficient $\omega_v(I,I)=\omega_v((R_{\pi/3})^3,(R_{\pi/3})^3)$. Note that a vison only acquires a trivial (+1) phase factor when it encircles any number of spinons six times. Therefore if $(R_{\pi/3})^6=-1$ for visons, there must be a nontrivial in-gap bound state (beyond spinon or vison) localized at the $\Omega=\pi/3$ disclination. Therefore the absence of in-gap bound state in the disclination indicates
\bea
\label{PSG:c6 no disclination}
(R_{\pi/3})^6=I^2=\bse\longrightarrow \omega_v(I,I)=1.
\eea
Notice that this argument only applies to a plaquette-centered rotation/inversion (such as $R_{\pi/3}$ and $I$ here), because the visons are located on plaquettes. For a site-centered crystal rotation, similar conclusions are not valid anymore.

To summarize, if a $Z_2$ SL does not host any gapless edge states protected by symmetries or in-gap disclination bound states, its vison PSG must satisfy conditions (\ref{PSG:T1 no edge})-(\ref{PSG:c6 no disclination}). All together this leads to the vison PSGs shown in TABLE \ref{tab:unification}, assuming the absence of symmetry-protected gapless edge states or in-gap disclination bound states.

\begin{table*}[tb]
\centering
\begin{tabular}{|c|c|c|c|c|c|c|c|c||c|c|}
\hline \multicolumn{9}{|c||}{Abrikosov fermion representation\cite{Lu2011} (\fsr)}&\multicolumn{2}{c|}{Schwinger boson representation\cite{Wang2006}(\bsr)}\\
\hline $\#$ & $\eta_{12}$&$\eta_\bss$&$\eta_{\bss\bst}$&$\eta_{\bss\cs}$&$\eta_{\cs\bst}$&$\eta_\cs$&Label&\multirow{2}{2cm}{Perturbatively gapped?}&$(p_1,p_2,p_3)$&Label\\
&&&&&&&&&&\\
\hline 1&+1&+1&+1&+1&+1&+1&$Z_2[0,0]A$&Yes&(1,1,0)&\\
\hline \alert{{2}}&-1&+1&+1&+1&+1&-1&\alert{${Z_2[0,\pi]\beta}$}&{Yes}&{{(0,1,0)}}&\alert{$Q_1=Q_2$ state}\\
\hline 5&+1&+1&+1&-1&-1&-1&$Z_2[0,0]B$&Yes&(1,0,1)&\\
\hline 6&-1&+1&+1&-1&-1&+1&$Z_2[0,\pi]\alpha$&No&(0,0,1)&$Q_1=-Q_2$ state\\
\hline 13&+1&-1&-1&-1&-1&-1&$Z_2[0,0]D$&Yes&(1,1,1)&\\
\hline 14&-1&-1&-1&-1&-1&+1&$Z_2[0,\pi]\gamma$&No&(0,1,1)&\\
\hline 15&+1&-1&-1&+1&+1&+1&$Z_2[0,0]C$&Yes&(1,0,0)&\\
\hline {{16}}&-1&-1&-1&+1&+1&-1&$Z_2[0,\pi]\delta$&No&(0,0,0)&\\
\hline
\end{tabular}
\caption{Correspondence between Schwinger boson mean-field (SBMF) states\cite{Wang2006} and Abrikosov fermion states\cite{Lu2011} of spin-1/2 symmetric \ztsl s on the kagome lattice. All 8 different SBMF (\bsr) states have their counterparts among 20 distinct Abrikosov fermion (\fsr) states. $Q_1=Q_2$ state\cite{Sachdev1992} in Schwinger boson representation is equivalent to $Z_2[0,\pi]\beta$ state\cite{Lu2011} in Abrikosov fermion representation, which is the only gapped \ztsl~in the neighbor of energetically favorable $U(1)$ Dirac state\cite{Ran2007}. On the other hand $Q_1=-Q_2$ state\cite{Sachdev1992} or $(0,0,1)$ state in Schwinger boson representation corresponds to $Z_2[0,\pi]\alpha$ state in Abrikosov fermion representation. ``Perturbatively gapped'' means that fermion spinons can have a fully-gapped mean-field spectrum by perturbing the nearest neighbor (NN) hopping ansatz.}
\label{tab:ansatz}
\end{table*}

\section{Unification of parton mean-field theories on the kagome lattice}\label{UNIFICATION}

With all results established in previous sections, now we are ready to establish the vison PSG $\{\omega_v(g,h)\}$ and fermionic spinon PSG $\{\omega_f(g,h)\}$ in any SBMF \ztsl~state. This allows us to unify \ztsl~states in Schwinger boson (\bsr) and Abrikosov fermion (\fsr) representations, as summarized in TABLE \ref{tab:ansatz}.  

In \bsr~or Schwinger boson construction\cite{Auerbach1994B}, a spin-$1/2$ on lattice site ${\bf r}$ is decomposed into two species of bosonic spinons $\{b_{{\bf r},\alpha}|\alpha=\uparrow/\downarrow\}$:
\bea
\vec{S}_{\bf r}=\frac12\sum_{\alpha,\beta=\uparrow/\downarrow}b^\dagger_{{\bf r},\alpha}\vec\sigma_{\alpha,\beta} b_{{\bf r},\beta}
\eea
where $\vec\sigma$ are Pauli matrices. Meanwhile in \fsr~or Abrikosov-fermion approach\cite{Abrikosov1965} spin-$1/2$ is represented by two flavors of fermionic spinons\cite{Affleck1988b} $\{f_{{\bf r},\alpha}|\alpha=\uparrow/\downarrow\}$
\bea
\vec{S}_{\bf r}=\frac12\sum_{\alpha,\beta=\uparrow/\downarrow}f^\dagger_{{\bf r},\alpha}\vec\sigma_{\alpha,\beta} f_{{\bf r},\beta}
\eea
To faithfully reproduce the 2-dimensional Hilbert space for spin-$1/2$, there is a \emph{single-occupancy constraint}: $\sum_\alpha b^\dagger_{{\bf r},\alpha}b_{{\bf r},\alpha}=\sum_\alpha f^\dagger_{{\bf r},\alpha}f_{{\bf r},\alpha}=1$ on every lattice site $\forall~{\bf r}$. The variational wavefunctions are obtained by implementing Gutzwiller projections\cite{Gros1989} on spinon mean-field ground state $|MF\rangle$, in order to enforce the single-occupancy constraint. Here $|MF\rangle$ is the ground state of \emph{(quadratic) mean-field ansatz} for bosonic spinons\cite{Sachdev1992,Wang2006}:
\bea
\hat{H}_{MF}^b=\sum_{{\bf x},{\bf y}}\sum_{\alpha,\beta}A_{{\bf x,y}}b^\dagger_{{\bf x},\alpha}b_{{\bf y},\alpha}+B_{{\bf x,y}}b_{{\bf x},\alpha}\epsilon^{\alpha\beta}b_{{\bf y},\beta}+~h.c.\notag
\eea
and similarly
\bea
\hat{H}_{MF}^f=\sum_{{\bf x},{\bf y}}\bpm f^\dagger_{{\bf x},\uparrow}\\f_{{\bf x},\downarrow}\epm^T\bpm t_{{\bf x,y}}&\Delta_{\bf x,y}\\ \Delta^\ast_{\bf x,y}&-t^\ast_{\bf x,y}\epm\bpm f_{{\bf y},\uparrow}\\f^\dagger_{{\bf y},\downarrow}\epm+~h.c.\notag
\eea
for fermionic spinons\cite{Wen1991c,Wen2002}. Proper on-site chemical potentials guarantee single-occupancy in $|MF\rangle$ on average.

The physical properties of a gapped $Z_2$ SL described by a projected wavefunction can be understood in terms of its mean-field ansatz. Specifically in \bsr~and \fsr~of $Z_2$ SLs, different PSGs for bosonic and fermionic spinons lead to distinct hopping/pairing patterns in mean-field ansatz $H^b_{MF}$ anf $H^f_{Mf}$. As pointed out in \Ref{Wang2006} there are 8 different Schwinger-boson (\bsr) mean-field ansatz of $Z_2$ SLs on kagome lattice, while 20 distinct mean-field ansatz of \ztsl~exists in \fsr~as shown in \Ref{Lu2011}. A natural question is: what is the relation between the $Z_2$ SLs in \bsr~and those in \fsr? Can they describe the same $Z_2$ SL phase or not?

To answer this question, we use their vison symmetry fractionalization pattern (or PSG) to determine the (in)equivalence of the two representations. To be precise, as discussed in (\ref{sym frac:z2 sl}), a Schwinger-boson ansatz corresponds to the same phase as an Abrikosov-fermion ansatz \emph{if and only if} they share the same vison PSG $\{\omega_v(g,h)\}$ and the same fermionic spinon PSG $\omega_f(g,h)$. As mentioned earlier, vison PSGs of a $Z_2$ SL can be probed by checking whether symmetry-protected edge states or in-gap disclination bound states exist or not. One important observation is that \emph{none of the SBMF states of $Z_2$ SLs constructed in Schwinger-boson approach supports any gapless edge state or in-gap disclination bound state}. This can be verified by computing the edge spectrum or defect spectrum in a Schwinger boson mean-field ansatz. Any gapped Schwinger-boson $Z_2$ SL ansatz can be tuned continuously to a limit that on-site chemical potential dominates over pairing/hopping terms, where it is clear no in-gap modes exist in edge/defect spectra. Therefore the vison PSG $\{\omega_v(g,h)\}$ in any Schwinger-boson ansatz is fully fixed as in TABLE \ref{tab:unification}. Amazingly this result (last column in TABLE \ref{tab:unification}) agrees with the vison PSG computed microscopically from Schwinger boson ansatz\cite{Huh2011}.

As discussed previously, symmetry fractionalization patterns $\{\omega_a(g,h)\}$ (or PSGs) of all three anyons $\{a=b,f,v\}$ in a \ztsl~are restricted by fusion rules \eqref{fusion rules}. They must satisfy relation (\ref{fusion rule:psg}) due to fusion rules, \ie the product of fermionic spinon PSG and vison PSG equals the bosonic spinon PSG, up to nontrivial twist factors discussed in section \ref{sec:nontrivial fusion}. These nontrivial twist factors take place in row 6-8 and 11-12 in TABLE \ref{tab:unification}, for algebraic relations $I^2=\bse$,~$R_{x,y}^2=\bse$ and $R_{x,y}^{-1}T^{-1}R_{x,y}T=\bse$. Therefore from TABLE \ref{tab:unification}, we can determine the correspondence between a Schwinger-boson ansatz and an Abrikosov-fermion ansatz, if they describes the same SET phase with the same PSGs for all three anyons. More precisely, fusion rule constraint (\ref{fusion rule:psg}) with associated twist factors leads to the following conditions:
\bea
&\notag\eta_{12}=(-1)^{p_1+1},\\
&\notag\eta_\bss=(-1)^{p_2+p_3+1},\\
&\notag\eta_{\bss\cs}=(-1)^{p_3},\\
&\notag\eta_{\bss\bst}=(-1)^{p_2+p_3+1},\\
&\notag\eta_\cs=(-1)^{p_1+p_3+1},\\
&\eta_{\cs\bst}=(-1)^{p_3}.\label{psg condition}
\eea

It turns out all 8 distinct gapped SBMF $Z_2$ SL states described by Schwinger-boson approach (\bsr) can also represented by Abrikosov-fermion approach (\fsr), as summarized in TABLE \ref{tab:ansatz}. Moreover, the most promising Schwinger-boson state for kagome Heisenberg model, \ie $Q_1=Q_2$ state\cite{Sachdev1992}, describes the same gapped symmetric \ztsl~phase as the most promising $Z_2[0,\pi]\beta$ state\cite{Lu2011} in Abrikosov-fermion approach. We notice that 4 of these 8 correspondences (the top 4 rows in TABLE \ref{tab:ansatz}) have been obtained in \Ref{Yang2012}.

\section{Candidate $Z_2$ spin liquids for kagome Heisenberg antiferromagnet}\label{sec:KHM candidate}


We have noted that there are 20 different Abrikosov-fermion mean-field (\fsr) states and 8 different SBMF (\bsr) states of symmetric spin-1/2 $Z_2$ spin liquids on the kagome lattice. Here we will argue that one state is the most promising candidate for kagome Heisenberg model from the following two criteria (i) energetics, as elaborated below; and (ii) the requirement that the phase to be connected to a $q=0$ magnetically ordered state via a continuous transition. This candidate is $Q_1=Q_2$ SBMF state in Schwinger boson representation, or equivalently $Z_2[0,\pi]\beta$ state in Abrikosov fermion representation.


Numerical studies on the kagome lattice Heisenberg antiferromagnet supplemented by a 2nd neighbor antiferromagnetic coupling ($J_2$) reveal that\cite{Yan2012a} on increasing $J_2$, the quantum spin liquid phase is initially further stabilized, before undergoing a transition\cite{Yan2012a} into a ${\bf q}=0$ magnetic order around $J_2\simeq0.2 J_1$. Since the correlation length increases as the transition is approached, it is likely to be second order. The $q=0$ magnetic order is a coplanar magnetic ordered state in which the three sublattices have spins aligned along three directions at 120 degrees to one another. The Schwinger-boson state that naturally accounts for this is the $Q_1=Q_2$ state in \Ref{Sachdev1992}, where under condensation of bosonic spinons the ${\bf q}=0$ magnetic order develops via a continuous transition in the O(4) universality class\cite{Chubukov1994a}.  Furthermore, variational permanent wavefunctions\cite{Tay2011} based on the $Q_1=Q_2$ state were shown to give very competitive energies in particular for small and positive $J_2$, establishing it as a contending state.

The $Z_ 2[0,\pi]\beta$ mean-field state\cite{Lu2011} in Abrikosov fermion representation (\fsr), also satisfies the desirable properties above. It can be thought as the $s$-wave paired superconductor of fermionic spinons near the energetically favorable $U(1)$ Dirac SL\cite{Ran2007}. The $s$-wave pairing opens up a gap at the Dirac point of the underlying $U(1)$ Dirac SL and this implies that the $Z_ 2[0,\pi]\beta$ state could be energetically competing with the $U(1)$ Dirac SL\cite{Lu2011}. Although variational wavefunctions that include pairing are often found to have higher energy\cite{Iqbal2011} than the underlying $U(1)$ Dirac SL, we note that this is a restricted class of states accessible via the parton construction, and a more complete search may land in the superconducting (\ie \ztsl) phase. For our purposes we will be content that it is proximate to the energetically favorable $U(1)$ spin liquid state. One can also describe a continuous transition from this ztsl~to the coplanar ${\bf q}=0$ magnetically ordered state, although the argument here is more involved than in the case of the Schwinger-boson representation. We make this argument\cite{Lu2011a} in two parts - first by ignoring the effects of the gauge field and recalling\cite{Grover2008} a seemingly unrelated transition between an s-wave superconductor and a quantum spin Hall phase of the fermionic partons. The latter spontaneously breaks the $SU(2)$ spin rotation symmetry down to $U(1)$ that defines the direction of the conserved spin component. On including gauge fluctuations one can argue that the quantum spin Hall phase is to be identified with the ${\bf q}=0$ magnetic order\cite{Lu2011a}.

Starting with the $U(1)$ Dirac spin liquid, the s-wave superconductor representing the \ztsl~is obtained by including a superconducting `mass' term that gaps the Dirac dispersion. Similarly, the quantum spin Hall phase is obtained by introducing a distinct `mass' term that also gaps out the Dirac nodes. There are three such mass terms indexed by the direction of the conserved spin in the quantum spin Hall state. All of them anti-commute with the superconducting mass term, which implies that a continuous transition is possible between these phases\cite{Grover2008}. On integrating out the fermions, the coefficients of the mass terms form an $O(5)$ vector of order parameters (real and imaginary parts of the pairing and the three quantum spin Hall mass terms), described by $O(5)$ non-linear sigma model with a Wess-Zumino-Witten (WZW) term\cite{Abanov2000,Tanaka2005,Senthil2006,Grover2008,Lu2011a}. The presence of the WZW term implies a continuous phase transition from the superconductor to the quantum spin hall state with a spontaneously chosen orientation. This is most readily seen by noting that skyrmions of the quantum spin Hall director carry charge 2, which when condensed leads to a superconductor with $SU(2)$ spin rotation symmetry \cite{Grover2008}.  Now, on including gauge couplings the superconductor is converted into the gapped $Z_2[0,\pi]\beta$ SL state. The quantum spin Hall state is a gapped insulator coupled to a compact $U(1)$ gauge field, which is expected to confine and lead to a conventional ordered state. This is seen to be the non-collinear magnetically ordered phase with the vector chirality at ${\bf q}=0$  (for details see Appendix \ref{appendix1}). The photon is identified as the additional Goldstone mode that appears since the ${\bf q}=0$ state completely breaks spin rotation symmetry. This further confirms the identification between $Q_1=Q_2$ Schwinger-boson state and $Z_2[0,\pi]\beta$ Abrikosov-fermion state, since they are both in proximity to the same magnetic order via a continuous phase transition.\\


\begin{table*}[tb]
\centering
\begin{tabular}{|c||c|c|c|c|c|c|c||c|c|c|||c|c|}
\hline $\#$ & $\eta_{12}$&$\eta_\bss$&$\eta_{\bss\bst}$&$\eta_{\bss\cs}$&$\eta_{\cs\bst}$&$\eta_\cs$&Label&\multirow{2}{2cm}{Perturbatively gapped?}&X-edge&Y-edge&$(p_1,p_2,p_3)$&Label\\
&&&&&&&&&&&&\\
\hline 1&+1&+1&+1&+1&+1&+1&$Z_2[0,0]A$&Yes&0&0&(1,1,0)&\\
\hline \alert{{2}}&-1&+1&+1&+1&+1&-1&\alert{${ Z_2[0,\pi]\beta}$}&{Yes}&0&0&{{(0,1,0)}}&\alert{$Q_1=Q_2$ state}\\
\hline 3&+1&+1&+1&-1&+1&-1&$Z_2[\pi,\pi]A$&No&0&allowed&&\\
\hline 4&-1&+1&+1&-1&+1&+1&$Z_2[\pi,0]A$&No&0&allowed&&\\
\hline 5&+1&+1&+1&-1&-1&-1&$Z_2[0,0]B$&Yes&0&0&(1,0,1)&\\
\hline 6&-1&+1&+1&-1&-1&+1&$Z_2[0,\pi]\alpha$&No&0&0&(0,0,1)&$Q_1=-Q_2$ state\\
\hline 7&+1&-1&+1&-1&+1&-1&-&-&allowed&0&&\\
\hline 8&-1&-1&+1&-1&+1&+1&-&-&allowed&0&&\\
\hline 9&+1&-1&+1&+1&+1&+1&-&-&allowed&allowed&&\\
\hline 10&-1&-1&+1&+1&+1&-1&-&-&allowed&allowed&&\\
\hline 11&+1&-1&+1&+1&-1&-1&-&-&allowed&0&&\\
\hline 12&-1&-1&+1&+1&-1&+1&-&-&allowed&0&&\\
\hline 13&+1&-1&-1&-1&-1&-1&$Z_2[0,0]D$&Yes&0&0&(1,1,1)&\\
\hline 14&-1&-1&-1&-1&-1&+1&$Z_2[0,\pi]\gamma$&No&0&0&(0,1,1)&\\
\hline 15&+1&-1&-1&+1&+1&+1&$Z_2[0,0]C$&Yes&0&0&(1,0,0)&\\
\hline {{16}}&-1&-1&-1&+1&+1&-1&$Z_2[0,\pi]\delta$&No&0&0&(0,0,0)&\\
\hline 17&+1&-1&-1&+1&+1&-1&$Z_2[\pi,\pi]B$&No&0&0&&\\
\hline 18&-1&-1&-1&+1&+1&+1&$Z_2[\pi,0]B$&No&0&0&&\\
\hline 19&+1&-1&-1&+1&-1&-1&$Z_2[\pi,\pi]C$&No&0&allowed&&\\
\hline 20&-1&-1&-1&+1&-1&+1&$Z_2[\pi,0]C$&No&0&allowed&&\\
\hline
\end{tabular}
\caption{\label{tab:MF_ansatz}20 different Abrikosov-fermion $Z_2$
SLs on a kagome lattice in the notation of \Ref{Lu2011}. Among them state $\#2$ or $Z_2[0,\pi]\beta$ state corresponds to the same phase as $Q_1=Q_2$ state\cite{Sachdev1992} in Schwinger-boson representation with\cite{Wang2006} $(p_1,p_2,p_3)=(0,1,0)$. Meanwhile $\#6$ or $Z_2[0,\pi]\alpha$ state belongs to the same phase as the so-called $Q_1=-Q_2$ state in Schwinger-boson representation with $(p_1,p_2,p_3)=(0,0,1)$. ``Perturbatively gapped'' means that fermion spinons can reach a fully-gapped superconducting ground state by perturbing the nearest neighbor (NN) hopping ansatz. ``0'' is a trivial topological index, indicating the absence of symmetry protected gapless modes on the edge. Among the 20 Abrikosov-fermion states, 6 may host protected edge states on X-edge, while 6 may support protected edge modes on Y-edge. None of the 8 Abrikosov-fermion states that have counterparts in Schwinger-boson representation can support gapless edge states.}
\end{table*}

\section{Fermionic $Z_2$ spin liquids with symmetry protected edge states}\label{sec:tsc}


In the Schwinger-boson representation, due to the absence of protected gapless edge states (or in-gap disclination bound states), the vison PSG is completely fixed. Therefore the PSG of bosonic spinons ($b$) fully determines the SET phase in any SBMF state. In other words, two SBMF ansatz correspond to the same $Z_2$ SL phase \emph{if and only if} they share the same bosonic-spinon PSG $\{\omega_b(g,h)\}$.

However this is not true in the Abrikosov-fermion representation: \ie two distinct \ztsl~phases can share the same fermionic-spinon PSG in their Abrikosov-fermion mean-field ansatz. This is due to the band topology\cite{Hasan2010,Hasan2011,Qi2011} in an Abrikosov-fermion mean-field ansatz, which can lead to symmetry protected edge states in a \ztsl, not captured by the fermionic-spinon PSG $\{\omega_f(g,h)\}$. In particular, certain fermionic-spinon PSGs allow for a nontrivial band topology, manifested by gapless edge states protected by mirror reflection symmetry in a topological superconductor of Abrikosov-fermions\cite{Chiu2013,Morimoto2013}.

\subsection{Reflection protected X-edge states}

There are two types of open edges in a cylinder geometry, \ie X-edge and Y-edge in FIG. \ref{RealUnitCell}. Other translational-symmetric open edge directions can be obtained by $\rcs$ rotations on these two prototype edges. As mentioned earlier, a cylinder with open X-edge preserves a symmetry group generated by $\{T,T_1,R_x\equiv(R_{\pi/3})^3R_y\}$ and $SU(2)$ spin rotations. It's straightforward to see that
\bea
\frac{\omega_f(T_1,T)}{\omega_f(T,T_1)}=\frac{\omega_f(T_1R_x,T_1R_x)}{\omega_f(R_x,R_x)}=1
\eea
for all Abrikosov-fermion states in TABLE \ref{tab:ansatz}. Since translation $T_1$ also commutes with spin rotations, it can be disentangled from other symmetries. As discussed in Appendix \ref{app:topo SC}, one can futher show that translational symmetry $T_1$ won't give rise to any nontrivial topological index. Focusing on time reversal $T$, mirror reflection $R_x$ (and $SU(2)$ spin rotations which commute with both $T$ and $R_x$), when acting on Abrikosov fermions $\{f_{{\bf r},\sigma}\}$ they satisfy
\bea
(U_{R_x})^2=\eta_\bss^{\hat F},~~~U_{R_x}^{-1}U_{T}^{-1}U_{R_x}^\ast U_T=\eta_{\bss\bst}^{\hat F}.
\eea
where $\hat F=\sum_{{\bf r},\sigma}f^\dagger_{{\bf r},\sigma}f_{{\bf r},\sigma}$ stands for the total fermion number. As shown in Appendix \ref{app:topo SC}, only when
\bea
\eta_{\bss}=-1,~~\eta_{\bss\bst}=+1.
\eea
will there be a nontrivial integer index $\mbz$ for topological superconductors. As summarized in TABLE \ref{tab:MF_ansatz}, there are 6 fermionic-spinon PSGs ($\#7\sim\#12$) that may support such a topological superconductor of fermonic spinons.

\subsection{Reflection protected Y-edge states}

On a cylinder with open Y-edge the symmetry group is generated by $\{T,T_y\equiv T_1^{-1}T_2^2,R_y\}$ and $SU(2)$ spin rotations. Again one can easily show that
\bea
\frac{\omega_f(T_y,T)}{\omega_f(T,T_y)}=\frac{\omega_f(T_yR_y,T_yR_y)}{\omega_f(R_y,R_y)}=1
\eea
in all Abrikosov-fermion states (see TABLE \ref{tab:ansatz}), and we can disentangle translation $T_y$ from other symmetries. Reflection $R_y$ and time reversal $T$ on Y-edge act on Abrikosov fermions with
\bea
(U_{R_y})^2=(\eta_\bss\eta_{\bss\cs})^{\hat F},~~~U_{R_y}^{-1}U_T^{-1}U_{R_y}^\ast U_T=(\eta_{\bss\bst}\eta_{\cs\bst})^{\hat F}.
\eea
Similarly a nontrivial integer index $\mbz$ for protected edge states can only happen when
\bea
\eta_\bss\eta_{\bss\cs}=-1,~~~\eta_{\bss\bst}\eta_{\cs\bst}=+1.
\eea
It turns out that only 6 fermionic-spinon PSGs ($\#3,\#4,\#9,\#10,\#19,\#20$) among all 20 cases in TABLE \ref{tab:MF_ansatz} may support topological superconductors with protected gapless modes on Y-edge, while the other 14 are not allowed.

\subsection{An example from minimal Dirac model}\label{subsec:topo SC edge}

As one example, we present a continuum model based on the minimal Dirac Hamiltonian for the mirror-protected topological superconductor in symmetry class CI\cite{Altland1997,Schnyder2008,Kitaev2009}. In the root state ($\nu=1$) which generates the integer ($\nu\in\mbz$) topological index, the low-energy edge excitations are described by 2 pairs of counter-propagating fermion modes (see Appendix \ref{app:topo SC} for derivations)
\beq
\mathcal{L}^0_{edge}=\sum_{a=\uparrow,\downarrow}\imth\psi^\dagger_{R,a}(\partial_t-\partial_x)\psi_{R,a}-\imth\psi^\dagger_{L,a}(\partial_t+\partial_x)\psi_{L,a}
\eeq
where the velocity is normalized as unity. The fermion modes transform under symmetries (time reversal $T$, reflection $R$ and spin rotations) in the following way:
\bea
&\psi_{\alpha,a}\overset{T}\longrightarrow\sum_{\beta,b}[\tau_x]_{\alpha,\beta}[\imth\sigma_y]_{a,b}\psi_{\beta,b}\\
&\psi_{\alpha,a}\overset{R}\longrightarrow\sum_{\beta,b}[\tau_x]_{\alpha,\beta}[\imth\sigma_y]_{a,b}\psi^\dagger_{\beta,b}\\
&\psi_{\alpha,a}\overset{\exp({\imth\theta\hat{n}\cdot\vec S})}\longrightarrow \sum_b\big[e^{\imth\frac{\theta}{2}\hat{n}\cdot\vec\sigma}\big]_{a,b}\psi_{\alpha,b}
\eea
where we use index $\alpha=R/L$ and $\vec\tau$ matrices for chirality (right/left movers), index $a=\uparrow/\downarrow$ and $\vec\sigma$ matrices for spin. It's straightforward to see that $TR=RT$ and $R^2=-1$. There are two kinds of backscattering terms between right and left movers, which preserve $SU(2)$ spin rotational symmetry:
\bea
&\mathcal{H}_{hop}=\sum_a\Big(t~\psi^\dagger_{R,a}\psi_{L,a}+t^\ast~\psi^\dagger_{L,a}\psi_{R,a}\Big),\notag\\
&\mathcal{H}_{pair}=\sum_{a,b}\Big(\Delta\psi_{R,a}[\imth\sigma_y]_{a,b}\psi_{L,b}+\Delta^\ast\psi^\dagger_{L,b}[\imth\sigma_y]_{a,b}\psi^\dagger_{R,a}\Big)\notag.
\eea
Among them, imaginary pairing is forbidden by time reversal $T$, while hopping and real pairing are both forbidden by mirror reflection $R$.

\section{Conclusion and outlook}\label{sec:conclusion}

In this work we systematically establish a connection between two different representations of $Z_2$ SLs, \ie Schwinger-boson representation (\bsr) and Abrikosov-fermion representation (\fsr). In the presence of physical symmetries, symmetry fractionalization patterns (manifested as projective symmetry groups, or PSGs in slave-particle/parton constructions) of anyons characterizes a symmetric \ztsl, or more generally a SET phase. We show that the vison PSGs can be determined by the absence of symmetry protected edge modes or in-gap bound states localized at crystal defect in $Z_2$ SLs. Observing that there are no symmetry-protected edge modes or defect bound states in any Schwinger-boson mean-field state, we show that vison PSG in any SBMF state is fully fixed. Utilizing the fusion rule constraint (\ref{fusion rule:psg}) relation between PSGs of three anyon types in a \ztsl, we obtain the fermion PSG in any SBMF state, and hence a correspondence between Schwinger-boson and Abrikosov-fermion states is achieved. 

Applying this general framework to kagome lattice $Z_2$ SLs, we showed that all 8 distinct Schwinger-boson (\bsr) mean-field states have their counterparts in Abrikosov-fermion (\fsr) representation. In particular we found that two energetically favorable states, Schwinger-boson $Q_1=Q_2$ state\cite{Sachdev1992} and Abrikosov-fermion $Z_2[0,\pi]\beta$ state\cite{Lu2011}, in fact belong to the same gapped symmetric \ztsl~phase in proximity to ${\bf q}=0$ non-collinear magnetic order. We argue that this phase is the most promising candidate for the observed $Z_2$ SL ground state in kagome Heisenberg model. 

With this connection in hand, we can potentially have a full understanding of the possible proximate phases and quantum phase transitions out of a \ztsl. It is well-known that the Schwinger-boson approach allows one to identify quantum phase transitions (QPT) into neighboring magnetic-ordered phases from a $Z_2$ SL, through condensation of bosonic spinons\cite{Sachdev1992}. Meanwhile, knowing the vison PSGs one can study possible QPTs between paramagnetic valence-bond-solid (VBS) phases and $Z_2$ SLs. On the other hand, in \fsr~it's straightforward to track down gapless spin liquid phases connected to a gapped $Z_2$ SL through a phase transition, as well as proximate superconducting ground states upon doping a quantum SL\cite{Lee2006}. Therefore the identification between \fsr~and \bsr~can point to a full phase diagram near a gapped $Z_2$ SL.

The correspondence obtained here also serves as important guidance towards a complete specification of symmetric $Z_2$ spin liquids on the kagome lattice. If one of the two promising states we identified is indeed the ground state of kagome Heisenberg model, then it provides a clear target for future studies to look for ``smoking gun'' signatures of these two states. Finally, we point out that similar studies can be applied to $Z_2$ SLs on the square lattice~\cite{Wang2011a,Jiang2012}, which is a direction for future works.

\acknowledgements
We thank M. Hermele, Y. Ran, C. Xu, Y.B Kim, T. Grover, M. Lawler, P. Hosur, F. Wang, S. White, T. Senthil, P.A. Lee, D.N Sheng and S. Sachdev for helpful discussions. We thank Mike Zaletel for penetrating comments and for collaborations on a related paper\cite{Zaletel2015}. GYC  thanks Joel E. Moore for support and encouragement during this work. YML is indebted to Shenghan Jiang for pointing out a typo in TABLE \ref{tab:unification} and for bringing \Ref{Yang2012} to our attention. The authors acknowledge support from Office of BES, Materials Sciences Division of the U.S. DOE under contract No. DE-AC02-05CH11231 (YML,AV), NSF DMR-1206515, DMR-1064319 and ICMT postdoctoral fellowship at UIUC (GYC) and in part from the National Science Foundation under Grant
No. PHYS-1066293(YML). YML thanks Aspen Center for Physics for hospitality where part of the work is finished. GYC is especially thankful to M. Punk, Y. Huh and S. Sachdev for bringing \Ref{Huh2011} to our attention and for the helpful discussions and suggestions.

\appendix

\section{Deriving vison PSGs in TABLE \ref{tab:unification} from edge states}\label{app:vison PSG}

In this section we explicitly show how to determine the vison PSGs in the last column of TABLE \ref{tab:unification}. First of all we can always choose a proper gauge by multiplying a proper $\pm1$ sign to symmetry actions $T_{1,2}$, so that
\bea
T_2\rcs=\rcs T_1,~~~T_1\rcs=\rcs T_2^{-1}T_1.
\eea
In other words both the 2nd and 3rd rows of TABLE \ref{tab:unification} are $+1$. Meanwhile as discussed in the end of section \ref{sec:edge->vison}, we have
\bea
T_1T_2=-T_2T_1.
\eea
for visons in a spin-$1/2$ $Z_2$ SL on kagome lattice.

The absence of symmetry protected edge states along X-edge leads to conditions (\ref{PSG:T1 no edge}). In particular we have
\bea
&T_1^{-1}T^{-1}T_1T=1,\notag\\
&R_x^2=(R_{\pi/3}R_y)^2=1\notag.
\eea
and
\bea
R_x^{-1}T^{-1}R_xT=(\rcs^{-1}T^{-1}\rcs T)\cdot(R_y^{-1}T^{-1}R_yT)=1\notag.
\eea
and
\bea
&\notag T_1R_x^{-1}T_1R_x=(T_1^{-1}T_2R_y^{-1}T_2R_y)\cdot(T_2^{-1}T_1^{-1}T_2T_1)\\
&=-T_1^{-1}T_2R_y^{-1}T_2R_y=1.\notag
\eea

At the same time, conditions (\ref{PSG:T1^(-1)T2^2 no edge}) come from the absence of protected edge states along Y-edge. Therefore we have
\bea
R_y^2=R_y^{-1}T^{-1}R_yT=1.
\eea
and
\bea
&\notag T_1^{-1}T_2^2R_y^{-1}T_1^{-1}T_2^2R_y=\\
&(T_1^{-1}R_y^{-1}T_1R_y)\cdot(T_2^{-1}T_1^{-1}T_2T_1)=-T_1^{-1}R_y^{-1}T_1R_y=1.\notag
\eea

These conditions fix all the vison PSGs except for $(\rcs)^6$ and $T_2^{-1}T^{-1}T_2T$. The latter one is easily determined as
\bea
T_2^{-1}T^{-1}T_2T=1.
\eea
by the absence of protected edge states in a cylinder whose edges are parallel to the direction of translation $T_2$. As discussed in section \ref{sec:edge->vison}, $(\rcs)^6=1$ is determined by the absence of protected mid-gap states in a disclination.

\section{Vison PSGs obtained by explicit calculations\cite{Huh2011}}

In this section we deduce the vison PSGs from the dual frustrated Ising model obtained in \Ref{Huh2011}, which describes vison fluctuations of Schwinger-boson $Z_2$ SLs on kagome lattice. The 4-component vison modes $\{v_n|1\leq n\leq4\}$ in section III~A of \Ref{Huh2011} transform under symmetry $g$ as
\bea
v_m\overset{g}\longrightarrow\sum_{n=1}^4~v_n\cdot \big[O_\phi(g)\big]_{n,m}
\eea
where the matrices $\{O_\phi(g)\}$ are given by
\bea
&O_{\phi}(T_{1}) =
-\left[
\begin{array}{cccc}
0& 0&0 &1 \\
0 & 0&-1 &0\\
0&1&0&0 \\
-1&0&0&0
\end{array}
\right],\\
&O_{\phi}(T_{2}) =
\left[
\begin{array}{cccc}
0& 0&-1&0 \\
0& 0&0 &-1\\
1&0&0&0 \\
0&1&0&0
\end{array}
\right],\\
&O_{\phi}(R_y) =
\left[
\begin{array}{cccc}
1& 0&0 &0 \\
0& 0&1 &0\\
0&1&0&0 \\
0&0&0&-1
\end{array}
\right],\\
&O_{\phi}(R_{\pi/3}) =
\left[
\begin{array}{cccc}
0& 0&1 &0 \\
1& 0&0 &0\\
0& 1&0&0 \\
0&0&0&1
\end{array}
\right].
\eea
Note that \Ref{Huh2011} considers mirror reflection $I_x=\rcs R_y$ with
\bea
O_{\phi}(I_x) =O_{\phi}(\rcs)O_{\phi}(R_y)=
\left[
\begin{array}{cccc}
0& 1&0 &0 \\
1& 0&0 &0\\
0&0&1&0 \\
0&0&0&-1
\end{array}
\right]\notag
\eea
It's straightforward to check the vison PSGs
\begin{align}\label{sym:w_vison}
&O_{\phi}(T_{1})O_{\phi}(T_{2})O_{\phi}(T_{1})^{-1}O_{\phi}(T_{2})^{-1} =-1; \nonumber \\
&O_{\phi}(R_{\pi/3})^{-1}O_{\phi}(T_{1})O_{\phi}(R_{\pi/3})O_{\phi}(T_{2}) =1; \nonumber \\
&O_{\phi}(R_{\pi/3})^{-1}O_{\phi}(T_{2})O_{\phi}(R_{\pi/3})O_{\phi}(T_{2})^{-1}O_{\phi}(T_{1})=1; \nonumber \\
&O_{\phi}(T_{1})O_{\phi}(T_{2})^{-1}O_{\phi}(R_{y})O_{\phi}(T_{2})^{-1}O_{\phi}(R_{y})^{-1}=-1; \nonumber \\
&O_{\phi}(T_{1})O_{\phi}(R_{y})O_{\phi}(T_{1})^{-1}O_{\phi}(R_{y})^{-1}=-1; \nonumber \\
&O_{\phi}(R_{\pi/3})^{6} = 1\nonumber\\
&O_{\phi}(R_{y})^{2} = 1\nonumber\\
&O_{\phi}(R_{\pi/3})O_{\phi}(R_{y})O_{\phi}(R_{\pi/3})O_{\phi}^{-1}(R_{y}) =1
\end{align}
which agree with the last column of TABLE \ref{tab:unification}.

\section{Two-dimensional TRI singlet superconductors (Class CI) with mirror reflection symmetry}\label{app:topo SC}

In this section we discuss possible symmetry-protected edge states of a time-reversal-invariant (TRI) singlet superconductor with mirror reflection symmetry in two dimensions. In a cylinder geometry, the symmetry group is generated by translation $T_e$ along the open edge (or cylinder circumference), mirror reflection $R$, time reversal $T$ and $SU(2)$ spin rotations. Without loss of generality, let's assume that $SU(2)$ spin rotations commute with all other symmetries $\{T,T_e,R\}$. We further assume that translation $T_e$ acts as
\bea\label{translation:trivial}
T_e^{-1}T^{-1}T_eT=T_eR^{-1}T_3R=+1.
\eea
for fermions.

\subsection{Classification}

Since translation $T_e$ has trivial commutation relation with other symmetry operations, it can be disentangled from the full symmetry group. Are there any gapless edge states protected by translation symmetry? This corresponds to ``weak index''\cite{Ran2010} of 2d topological superconductors in class CI (with time reversal and $SU(2)$ spin rotations), which is nothing but 1d topological index of the same symmetry class. Class CI has trivial classification (0) in 1d, therefore we don't have translation-protected edge states. Due to absence of 2d topological index in class CI, any protected edge states must come from mirror reflection symmetry $R$. The classification of mirror reflection protected topological insulators/superconductors is resolved in \Ref{Chiu2013,Morimoto2013} in the framework of K-theory\cite{Kitaev2009}. The classification of non-interacting topological phases of fermions in class CI with mirror reflection $R$ depends on the commutation relation
\bea
R^2=s_1,~~~R^{-1}T^{-1}RT=s_2;~~~s_i=\pm1.
\eea
with time reversal $T$. When $s_1=s_2=+1$, the K-theory classification is given by $\pi_0(R_6)=0$, \ie no topological superconductors with protected edge states. When $s_1=s_2=-1$, the classification is $\pi_0(C_5)=0$ \ie no topological superconductors. When $s_1=+1,s_2=-1$ the classification is $\big[\pi_0(R_5)\big]^2=0^2=0$ \ie no topological superconductors. Only when
\bea
s_1=-1,~~~s_2=+1.
\eea
the classification is given by $\pi_0(R_4)=\mbz$ and there are topological superconductors with an integer index ($\mbz$).

So far we've only considered global symmetries together with spatial mirror reflections to arrive at this integer $\mbz$ classification for non-interacting electrons. In a symmetric spin liquid of spin-$1/2$ particles on a kagome lattice, all space group symmetries need to to taken into account, and the interactions between fermionic spinons are also important. Other space group symmetries and the single-occupancy constraints for Abrikosov fermions will impose extra conditions and reduce this integer classification. Interaction effects may further reduce the classification. Therefore in TABLE \ref{tab:MF_ansatz} we've denoted the 6 states (for X- and Y-edge each) as ``allowed'' to support protected gapless edge states.

\subsection{Minimal Dirac model and protected edge states}

Here we construct a Dirac model for the root state ($\nu=1$) of topological superconductors with integer index $\nu\in\mbz$ and mirror reflection $R$ satisfying
\bea
R^2=-1,~~~RT=TR.
\eea
Writing spin-$1/2$ electrons in the Nambu basis $\psi_k\equiv(c_{k,\uparrow},c^\dagger_{-k,\downarrow})^T$, we use Pauli matrices $\vec\tau$ for Nambu index and $\vec\mu,\vec\rho$ for orbital index. The 8-band massless Dirac Hamiltonian is given by
\bea
H_{Dirac}=\sum_k(k_x\mu_x+k_y\mu_z)\rho_y\tau_x
\eea
The Dirac fermion transforms as
\bea
&\notag\psi_k\overset{T}\longrightarrow\imth\tau_y\psi_{k}^\ast,\\
&\psi_{(k_x,k_y)}\overset{R}\longrightarrow\imth\mu_x\rho_y\psi_{(k_x,-k_y)}
\eea
The only symmetry-allowed mass term is
\bea
M=\tau_z
\eea

Now let's create a mass domain wall across an open edge at $x=0$ along $y$-axis. The gapless edge states is captured by zero-energy solution of differential equation
\bea
-\imth\partial_x\mu_x\rho_y\tau_x+m(x)\tau_z=0,~~~m(x)=|m(x)|\cdot\text{Sgn}(x)\notag
\eea
which is $|edge\rangle\sim e^{-\int_0^xm(\lambda)\text{d}\lambda}|\mu_x\rho_y\tau_y=+1\rangle$. Therefore $\rho_y=\mu_x\tau_y$ for the protected gapless edge modes, localized on the edge between the root topological superconductor and the vacuum. The Hamiltonian for the protected edge states is given by
\bea
H_k=k_y\mu_z\rho_y\tau_x\rightarrow-k_y\mu_y\tau_z
\eea
Apparently there are 4 gapless modes: 2 right movers and 2 left movers. It's straightforward to simplify such edge states to the form in section \ref{subsec:topo SC edge}.

\section{$Z_ 2[0,\pi]\beta$ mean-field state and its proximate phases}\label{appendix1}

We begin by revisiting $U(1)$ Dirac spin liquid \cite{Hastings2000,Ran2007,Hermele2008} and $Z_{2}[0,\pi]\beta$ state\cite{Lu2011} and on a kagome lattice. The spin liquid states are proposed to be the ground state of nearest-neighbor spin-$1/2$ Heisenberg model on kagome lattice:
\beq
H = J \sum_{<ij>}{\hat S}_{i}\cdot {\hat S}_{j}
\eeq
The low-energy theory of the $U(1)$ Dirac spin liquid \cite{Hermele2008} is described by a 8-component spinor $\psi$ of fermionic spinons in Dirac spectrum and a strongly fluctuating $U(1)$ gauge field $a_{\mu}$:
\beq
H =\sum_{\bf k} \psi^{\dagger}_{{\bf k}}\sigma^{0}\mu^{0}(\tau_{x}k_{x} + \tau_{y}k_{y})\psi_{{\bf k}}.
\label{ASL}
\eeq
where $\tau^\nu$, $\sigma^{\nu}$ and $\mu^{\nu}$ are the Pauli matrices acting on Dirac, spin and valley indices of $\psi_{{\bf k}}$ (there are two nodes or two ``valleys" : hence $\psi_{{\bf k}}$ is a $8$-component spinor). Here we temporarily ignore the compact $U(1)$ gauge theory for clarity of discussion.

To obtain a $Z_{2}$ spin liquid from this $U(1)$ Dirac spin liquid, a BCS-type pairing term is introduced for the fermion $\psi$. We require the pairing term to be invariant under spin rotational symmetry, lattice symmetry operations (see Fig.\ref{RealUnitCell} for the symmetries of kagome lattice) and time-reversal symmetry, because the spin liquid found in DMRG study does not break any of the symmetries. Furthermore, the pairing should gap out the Dirac spectrum as the $Z_{2}$ spin liquid found numerically is fully gapped.
\beq
\delta H = \Delta \psi^{\dagger} (\sigma^{y}\mu^{y}\tau^{y})\psi^{*} + h.c.,
\label{pairing}
\eeq
which is a singlet of spin, valley and Dirac spinor indices. Being a singlet under spin and valley indices gaurantees that the pairing is invariant under spin rotational symmetry and lattice symmetry operations. The low-energy physics of $Z_{2}[0,\pi]\beta$ state is described by $H + \delta H$ in Eq.\eqref{ASL} and Eq.\eqref{pairing}, \ie a gapped singlet superconductor of fermionic spinons coupled to a dynamical $Z_2$ gauge field.

It is known that Dirac fermions are unstable (with sufficiently large interactions) to open up gap in various channels. Each channel is called a ``mass'' and is represented by a constant matrix in the Dirac spinor representation. For example in Dirac Hamiltonian (\ref{ASL}), $\tau^z\mu^\alpha\sigma^\beta$ are all mass terms, with $\alpha,\beta=0,x,y,z$.  For a 2+1-D Dirac fermion, when we can find five such mass matrices anti-commuting with each other\cite{Abanov2000,Senthil2006,Ryu2009}, we obtain a non-linear sigma model supplemented with a topological WZW term\cite{Wess1971,Witten1983} after integrating out {\it massive} fermions. This non-linear sigma model describes fluctuating order parameters of the Dirac fermions. Most importantly, the theory can describe a Landau-forbidden second-order transition between the two phases\cite{Senthil2004a}, where the transition is driven by condensing the topological defects\cite{Tanaka2005,Senthil2006,Grover2008,Lu2011a}. The mass terms associated to the $Z_2$ spin liquid are real and imaginary parings in (\ref{pairing}).

Because we seek for the nearby phases of the $Z_{2}$ spin liquid, the relevant mass terms should anti-commute with the pairing term \eqref{pairing} and the kinetic term \eqref{ASL}. We immediately find two $O(3)$ vector mass terms among $26$ mass terms of the $U(1)$ Dirac spin liquid\cite{Hermele2008}, anti-commuting with the pairing term \eqref{pairing}.

Among the two $O(3)$ vector mass terms, we consider only the $O(3)$ vector chirality operator\cite{Hermele2008} ${\hat V} \sim <\psi^{\dagger} \tau^{z}{\vec \sigma}\psi>$ to examine the magnetically ordered proximate phase of the  $Z_{2}[0,\pi]\beta$ state. The operator ${\hat V}$ is spin-triplet and time-reversal symmetric. This order parameter represents the sum of the vector chirality around honeycomb plaquette $H$ on Kagome lattice.
\beq
{\hat V}^{a} \sim \sum_{<ij> \in H} ({\vec S_{i}} \times {\vec S_{j}})^{a},
\eeq
Because the vector chirality ${\hat V}$ is spin-triplet, we expect the non-linear sigma model for the unit $O(5)$ vector $= (\Delta_{x},\Delta_{y}, {\hat V})$ to describe the transition between the spin liquid and a magnetically ordered phase with the non-zero $\langle {\hat V} \rangle$.

To see this, we approach the critical point between the spin liquid and the magnetically ordered phase from the ordered phase. The low-energy effective theory for the symmetry-broken phase, including the compact $U(1)$ gauge field $a_{\mu}$, is
\begin{align}
\mathcal{L} =  \psi^{\dagger}&\sigma^{0}\mu^{0}\tau^{\mu}\cdot (\imth\partial_{\mu}+a_{\mu})\psi + m{\hat V} \cdot \psi^{\dagger} \tau^{z}{\vec \sigma}\psi \nonumber\\
& + \frac{1}{g^{2}} (\partial_{\mu} {\vec V})^{2} + \frac{1}{2{\tilde e}^{2}}(\varepsilon^{\mu\nu\lambda}\partial_{\nu}a_{\lambda})^{2} +\cdots
\label{magnetic1}
\end{align}
In the symmetry-broken phase, the $O(3)$ vector order parameter ${\hat V}$ develops a finite expectation value, and we assume $\langle{\hat V}\rangle = (0,0, 1)$ without losing generality (other direction of $\langle{\hat V}\rangle$ can be generated by the spin rotations). With the expectation value ${\hat V}$, it is not difficult to see that the spin-up fermions and the spin-down fermions have an energy gap $|m|$ at the Dirac points with the opposite sign, and the mass gap consequently generates the ``spin Hall effect'' for the fermions. The quantum spin Hall effect has an important implication\cite{Ran2008a} on the fate of the compact gauge field $a_{\mu}$: it ties the gauge fluctuation to the spin fluctuation, and thus the Goldstone mode ($\sim$ spin fluctuation) of the spin ordered phase becomes a photon ($\sim$ gauge fluctuation) of $a_{\mu}$. This implies that the gauge field $a_{\mu}$ is in the Coulomb phase and the photon of $a_{\mu}$ is free to propagate. Hence there are three Goldstone modes in the magnetically ordered phase, one photon mode from the non-compact $U(1)$ guage field $a_{\mu}$ and two Goldstone modes from the ordering of the $O(3)$ vector ${\hat V}$. Meanwhile accompanying the proliferation of $a_\mu$ photons, fermionic spinons will be confined \cite{Polyakov1977} due to instanton effect of 2+1-D $U(1)$ gauge theory. Therefore indeed it is a non-collinear magnetic ordered phase with three Goldstone modes, which does not support fractionalized excitations.

Upon integrating out the massive Dirac fermion, we obtain the effective theory\cite{Abanov2000,Lu2011a} for the fluctuating ${\hat V}$ in the presence of the gauge field $a_{\mu}$
\beq
 \mathcal{L}= \frac{1}{g^{2}} (\partial_{\mu} {\hat V})^{2} + 2a_{\mu} J^{\mu}_{skyr} +  \frac{1}{2{\tilde e}^{2}}(\varepsilon^{\mu\nu\lambda}\partial_{\nu}a_{\lambda})^{2} +\cdots
\label{magnetic2}
\eeq
where $J^{\mu}_{skyr}$ is the skyrmion current of ${\hat V}$, e.g. $J^{0}_{skyr} \propto {\hat V}\cdot (\partial_{x} {\hat V}\times\partial_{y} {\hat V})$ is the skyrmion density of ${\hat V}$. From the coupling between $J^{\mu}_{skyr}$ and $a_{\mu}$, it is clear that the skyrmion carries the charge-2 of the gauge field $a_{\mu}$. Hence, condensing the skyrmion of ${\hat V}$ breaks $U(1)$ gauge group down to $Z_{2}$ and the skyrmion can be thought as the pairing $\sim \langle\psi^{\dagger}\psi^{\dagger}\rangle$ of the fermionic spinons $\psi$ in \eqref{ASL}. As the condensation of the skyrmion would destroy the ordering in ${\hat V}$ and induce the pairing between the fermionic spinons, we will enter the $Z_{2}$ spin liquid phase next to the symmetry-broken phase, \ie $Z_{2}[0,\pi]\beta$ state.

Thus we have established that the magnetically ordered phase next to the $Z_{2}[0,\pi]\beta$ state is a non-collinear magnetically ordered phase with the non-zero vector chirality at ${\bf q}=0$. Given that the ${\bf q}=0$ magnetically ordered state is also a non-collinear magnetically ordered phase with the non-zero vector chirality at ${\bf q}=0$, the $Z_{2}[0,\pi]\beta$ state is a natural candidate for the $Z_2$ SL proximate to the ${\bf q}=0$ magnetically ordered state.


\end{document}